\documentclass[pdflatex,sn-nature]{sn-jnl}

\usepackage{graphicx}%
\usepackage{multirow}%
\usepackage{amsmath,amssymb,amsfonts}%
\usepackage{amsthm}%
\usepackage{mathrsfs}%
\usepackage[title]{appendix}%
\usepackage{xcolor}%
\usepackage{textcomp}%
\usepackage{manyfoot}%
\usepackage{booktabs}%
\usepackage{algorithm}%
\usepackage{algorithmicx}%
\usepackage{algpseudocode}%
\usepackage{listings}%
\usepackage{float}
\usepackage{adjustbox}
\usepackage[left]{lineno}
\usepackage{natbib}
\usepackage{url}

\theoremstyle{thmstyleone}%
\theoremstyle{thmstyletwo}%

\theoremstyle{thmstylethree}%

\unnumbered

\begin{document}

\title[The Uncertain Policy Price of Scaling Direct Air Capture]{The Uncertain Policy Price of Scaling Direct Air Capture}


\author*[1,2,3]{\fnm{Leonardo} \sur{Chiani}\email{leonardo.chiani@polimi.it}} 
\author[1,2,3]{\fnm{Pietro} \sur{Andreoni}}
\author[2,3]{\fnm{Laurent} \sur{Drouet}}
\author[4]{\fnm{Tobias} \sur{Schmidt}}
\author[4,5]{\fnm{Katrin} \sur{Sievert}}
\author[4,5,6]{\fnm{Bjarne} \sur{Steffen}}
\author[1,2,3]{\fnm{Massimo} \sur{Tavoni}}
\affil*[1]{\footnotesize Department of Management Engineering, Politecnico di Milano, Italy}
\affil[2]{\footnotesize CMCC Foundation - Euro-Mediterranean Center on Climate Change, Italy}
\affil[3]{\footnotesize RFF-CMCC European Institute on Economics and the Environment, Italy}
\affil[4]{\footnotesize Energy and Technology Policy Group, ETH Zurich, Switzerland}
\affil[5]{\footnotesize Climate Finance and Policy Group, ETH Zurich, Switzerland}
\affil[6]{\footnotesize Center for Energy and Environmental Policy Research, Massachusetts Institute of Technology, USA}

\abstract{Direct air carbon capture and storage (DACCS) is a promising CO\textsubscript{2} removal technology, but its deployment at scale remains speculative. Yet, its technological, economic, and policy-related uncertainties have often been overlooked in mitigation pathways. This paper conducts the first uncertainty quantification and global sensitivity analysis of DACCS on technological, market, financial and public support drivers, using a detailed-process Integrated Assessment Model and newly developed sensitivity algorithms. We find that DACCS deployment exhibits a fat-tailed distribution: most scenarios show modest technology uptake, but there is a small but non-zero probability (4–6\%) of achieving gigaton-scale removals by mid-century. Scaling DACCS to gigaton levels requires subsidies that always exceed 200-330 USD/tCO\textsubscript{2} and are sustained for decades, resulting in a public support programme of 900-3000 USD Billions. Such an effort pays back by mid-century, but only if accompanied by strong emission reduction policies. These findings highlight the critical role of climate policies in enabling a robust and economically sustainable CO\textsubscript{2} removal strategy.}

\keywords{Direct Air Capture, Uncertainty Quantification, Global Sensitivity Analsysis, Mitigation}



\maketitle

\section{Introduction}
\label{sec:intro}

In the last decade, the Paris Agreement and announcements of net-zero pledges by many countries have increased the interest in deep mitigation pathways to limit global warming to well below 2°C. These pathways assume a transition to low-carbon energy sources combined with the use of carbon dioxide removal (CDR) technologies, essential for balancing emissions from 'hard-to-abate' sectors and absorbing excess CO\textsubscript{2} in the atmosphere \citep{IPCC_AR6_WGIII_2023}. However, CDR remains a subject of intense debate in the general public and the scientific community \citep{cox2020public, baum2024public}. This debate is centred around the deterrence these technologies may pose to mitigation \citep{grantConfrontingMitigationDeterrence2021, grantPolicyImplicationsUncertain2021}, inequality repercussions \citep{hickel2022existing, andreoniInequalityRepercussionsFinancing2024}, and removal costs \citep{sievertConsideringTechnologyCharacteristics2024, youngCostDirectAir2023}. 

As of 2024, CDR methods are estimated to remove roughly 2 GtCO\textsubscript{2} per year, primarily (99.9\%) in the land use, land-use change and forestry sector \citep{lamb2024carbon}. Accounting only for the 0.1\%, novel CDR technologies such as direct air carbon capture and storage (DACCS) and bioenergy with carbon capture and storage (BECCS) are still in their infancy, and the scale of their future deployment is largely uncertain \citep{lamb2024carbon}. DACCS has several advantages compared to other CDR methods: it allows for permanent removal of CO\textsubscript{2}, is less land-intensive, and enables straightforward accounting of removed emissions \citep{dufour2025avoid}. Relevant works highlight its potential in stringent mitigation scenarios \citep{realmonteIntermodelAssessmentRole2019, fuhrman2021role, strefler2021carbon, fuhrman2023diverse}. However, as a new technology, significant uncertainties exist regarding the barriers to scaling DACCS at the levels needed to make a difference, and failing to account for them may significantly delay the ecological transition and generate stranded assets \citep{bindl2025risks}.

These uncertainties are both directly and indirectly related to the technology itself. Direct uncertainties concern the costs and performance of different DACCS technologies \citep{shayeghFutureProspectsDirect2021, youngCostDirectAir2023, sievertConsideringTechnologyCharacteristics2024, abeggExpertInsightsFuture2024, wei2025unlocking}, their dependence on regional climatic conditions \citep{sendi2022geospatial, an2022impact}, and the scale of market penetration (also called feasible deployment) \citep{shayeghFutureProspectsDirect2021, grantPolicyImplicationsUncertain2021, edwardsModelingDirectAir2024, abeggExpertInsightsFuture2024}. Indirect uncertainties stem mainly from the policy environment, shaped by emissions reduction ambitions, targeted support for DACCS, and interactions with other CDR options. This policy dependence is critical \citep{sovacool2022climate}: DACCS has few co-benefits compared with alternatives such as BECCS or afforestation, and the current need for CO\textsubscript{2} removal is limited. No countries currently include novel CDR methods in their nationally determined commitments \citep{lamb2024carbon}, and for many sectors, reducing emissions remains cheaper than capturing and storing them with DACCS \citep{dufour2025avoid}.

Different computational tools, such as integrated assessment models (IAMs), were developed or expanded to help policymakers navigate these uncertainties. Yet, most studies examined them in isolation rather than in an integrated fashion. Some analyzed only the uncertainty in feasible deployment, without accounting for other factors \citep{grantConfrontingMitigationDeterrence2021, edwardsModelingDirectAir2024}. Others explored socio-economic pathways and policy settings while keeping other inputs, such as costs or deployment limits, fixed \citep{fuhrman2021role, fuhrman2023diverse}. Where sensitivity analysis had been applied, it often relied on one-at-a-time changes in uncertain inputs \citep{realmonteIntermodelAssessmentRole2019, hanna2021emergency}, covering only a small portion of the possible inputs' space and overlooking interactions between variables \citep{saltelli2010avoid}. A notable exception is a recent study that implemented multiple CDR technologies in a mixed-integer linear optimization model to design robust portfolios and assess the relative importance of different inputs using explainable machine learning indices \citep{mendez2025deep}. Still, this work left -- highly critical -- policy uncertainty unaddressed and represented technological uncertainty with only a few parameters. More generally, the existing literature shows a fragmentation due to the extreme complexity of the problem.

\begin{figure}[H]
    \centering
    \includegraphics[width=1\linewidth]{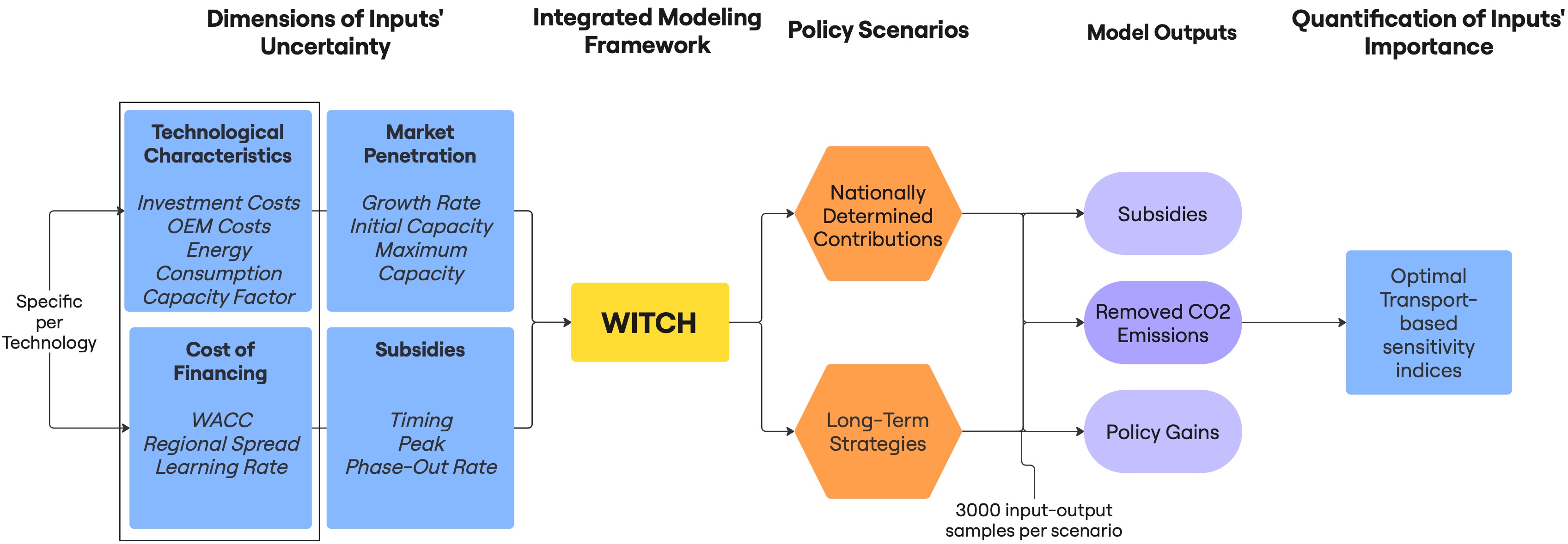}
    \caption{The study uses a probabilistic approach to explore four key dimensions of uncertainty: technological characteristics, market penetration, cost of financing, and subsidies (see Supplementary Table 1). These uncertainties are integrated within the WITCH integrated assessment model. Superimposed on the probabilistic structure, two baseline scenarios are considered: Nationally Determined Contributions and Long-Term Strategies. The model produces three key outputs: net CO\textsubscript{2} removals, subsidies, and policy economic gains/costs, which are then analysed using Optimal Transport-based sensitivity indices and other statistical methods.}
    \label{fig:workflow}
\end{figure}

Building on previous works, we address this issue by implementing three DACCS technologies in the detailed-process integrated assessment model (IAM) WITCH \citep{emmerling2016witch}: liquid solvent, solid sorbent, and CaO ambient weathering. We characterize the uncertainties around DACCS deployment relying on probability distributions that reflect the most up-to-date understanding of the uncertain inputs along all critical dimensions (technological characteristics, market growth, cost of financing, and public subsidies), and implement two scenarios with differentiated climate ambition. We then carry out the first global sensitivity analysis (GSA) of DACCS using advanced sensitivity analysis techniques based on Optimal Transport \citep{wiesel2022, borgonovo2024global}. The umbrella term GSA describes a set of tools to investigate how the variation in the output of a model can be attributed to variation of its inputs \citep{saltelli2002sensitivity, pianosi2016sensitivity}. In other words, GSA is a way to systematically address the role of numerical assumptions (the inputs) in supporting the decision-making process. In the context of climate mitigation research, GSA was used to explore robust abatement pathways \citep{lamontagne2019robust}, the importance of socio-economic assumptions \citep{marangoni2017sensitivity}, and technological characteristics \citep{bosettiSensitivityEnergyTechnology2015}, among others. GSA is a critical tool for understanding the behaviour of computational models, especially when those models inform high-stakes policy decisions \citep{saltelli2020five}. Deploying novel Carbon Dioxide Removal (CDR) technologies such as DACCS falls within this category. On the one side, these technologies are characterized by substantial uncertainties across technical, economic, and policy dimensions. On the other side, they hold the potential for notable societal gains or losses. Incorporating CDR into models without robust, multidimensional uncertainty analysis poses significant risks. Figure \ref{fig:workflow} summarizes the adopted workflow.

\section{Sensitivity Analysis for Technology Deployment}
\label{sec:satd}

We use the newly developed Optimal Transport-based sensitivity indices \citep{wiesel2022, borgonovo2024global} (OT-based indices, henceforth) as a GSA tool. This method builds on the mathematical framework of Optimal Transport \citep{villani2009, peyre2019computational}, a field at the intersection of probability, statistics, and optimization.
The OT-based indices offer several advantages that make them particularly well-suited to our application \citep{chiani2025global}. First, they are well-defined for multivariate outputs, like vectors and time-dependent curves. This property is practical when studying the deployment of new technologies, for which we are often interested in temporal dynamics. Second, unlike classical variance-based techniques, these indices remain easily interpretable when inputs are statistically dependent, allowing us to address realistic correlation structures among uncertain inputs. This property is essential in optimization models, where unfeasibilities may induce correlation structures in the inputs, even if they are independent in sampling. Third, their normalized $[0,1]$ scale makes interpretation intuitive, and they can be estimated directly from Monte Carlo samples, avoiding the need for additional model runs or surrogate modelling.

Defining the input distributions is a crucial step in the GSA process. In this study, we examine 36 input variables across four key dimensions of uncertainty (cf. Figure \ref{fig:workflow}): technological characteristics, market growth, cost of financing, and public subsidies. The technological characteristics are technology-specific and calibrated using a probabilistic DACCS cost model \citep{sievertConsideringTechnologyCharacteristics2024}. This category includes inputs such as capital expenditures, operational and maintenance costs, and energy requirements. To represent market growth, we constrain the expansion of DACCS deployment over time using a logistic growth function. This constraint ensures that investment decisions in each period are limited by the scale already achieved in previous periods. Market penetration input distributions are based on data from other technologies \cite{edwardsModelingDirectAir2024} and on ad-hoc assumptions. The cost of financing follows a recent implementation \citep{calcaterra2024reducing}, with weighted average cost of capital values centred on estimates from the literature \citep{fasihiTechnoeconomicAssessmentCO22019} and sampled from truncated Gaussian distributions to capture the diversity of financing costs across technology, country and time. Finally, public subsidies are represented as a normative policy input, specified through a simple parametric function defining their timing, peak and phase out rate. Supplementary Table 1 summarises the input distributions, with further implementation details described in Section \ref{sec:methods}.

We complement this probabilistic analysis with a traditional scenario-based one to address the deep and unquantifiable uncertainty in international climate policy. We analyze the deployment of DACCS under two baseline scenarios to capture distinct policy trajectories and ambition levels:
\begin{itemize}
    \item \textbf{Nationally Determined Contributions (NDC)}: This scenario extrapolates the current short-term policy commitments from countries under the Paris Agreement. The NDC scenario assumes the continuation of existing climate policies without significant additional ambition, reflecting the current heterogeneous and moderate global policy landscape. This scenario is a benchmark of realistic yet unambitious international climate policy commitments.
  \item \textbf{Long-Term Strategies (LTS)}: The LTS scenario represents an ambitious policy trajectory consistent with stringent global climate goals, where all major economies implement enhanced policies aligned with net-zero targets by mid-century. It reflects the net-zero commitments announced by several countries, leading to significant emission reductions consistent with the Paris agreement \citep{ALELUIAREIS2023105933}.
\end{itemize}
We draw 3000 input samples from the distributions in Supplementary Table 1 and run the model for each input sample in each scenario. Thus, our dataset of input-output realizations amounts to 6000 samples. All runs are feasible in the NDC scenario, while, in the LTS scenario, 417 runs over the 3000 total are infeasible, and we discard them.

\begin{figure}[H]
    \centering
    \includegraphics[width=1\linewidth]{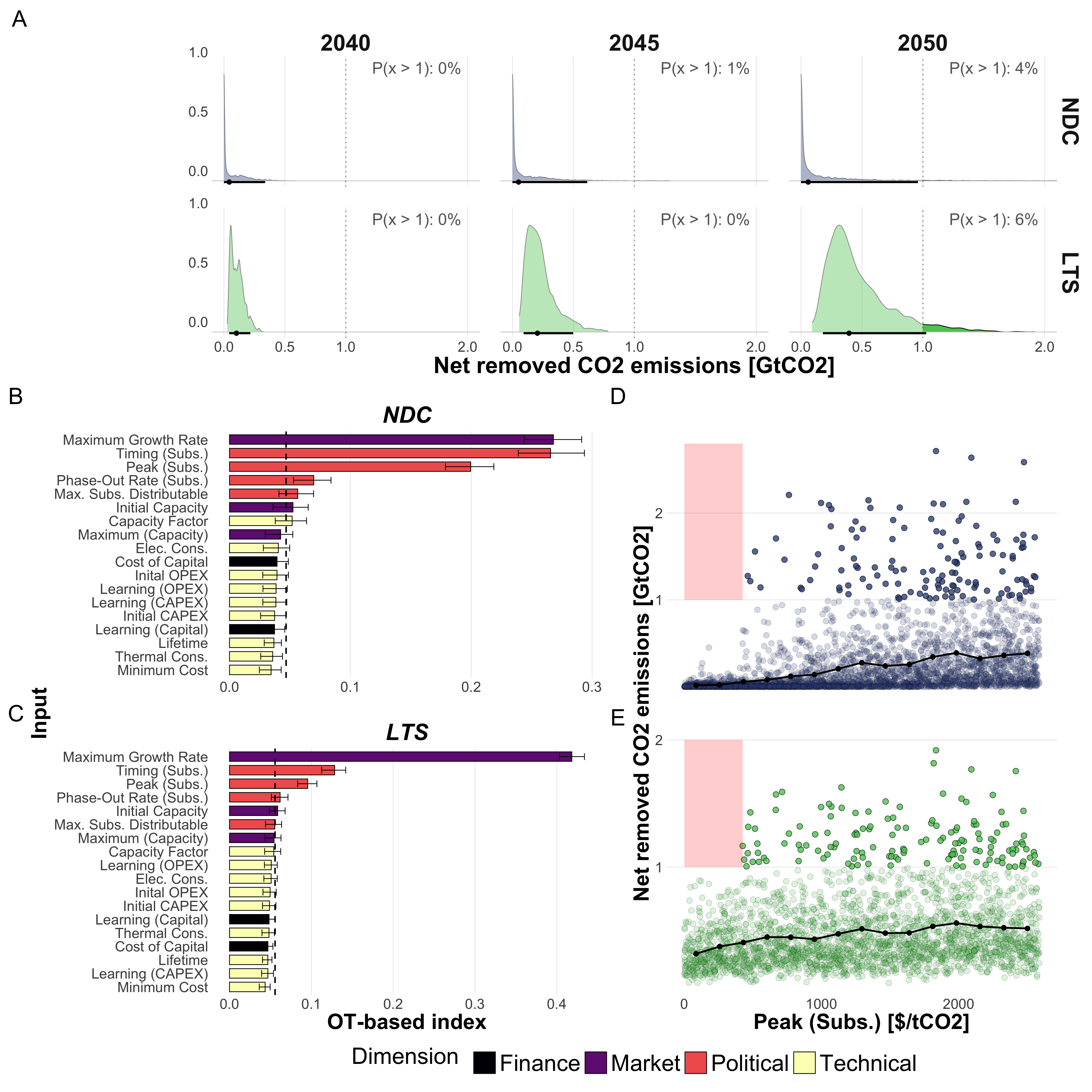}
    \caption{A. Yearly net CO\textsubscript{2} emissions removed by DACCS between 2040 and 2050 in the two scenarios (NDC and LTS). Each plot represents the probability density. Transparency identifies the threshold of 1 GtCO\textsubscript{2}. The black bars below the plots represent medians and 5th-95th quantile ranges. B-C. Sensitivity analyses for net removed CO\textsubscript{2} emissions in the two scenarios. The x-axis represents the OT-based sensitivity index. On the y-axis, each bar represents an input. Where inputs are differentiated by technology, we consider the maximum index among them. Error bars are the 95\% bootstrapped confidence intervals. Colours represent the dimension of uncertainty. Inputs are ranked in descending order. The dashed vertical line represents the irrelevance threshold. D-E. Partial dependency plots of the peak subsidies against the removed emissions in the two scenarios. The x-axis represents the value of the peak subsidies, and the y-axis represents the corresponding 2050 net removed CO\textsubscript{2} emissions in the two scenarios. The transparency is the 1 GtCO\textsubscript{2} threshold, the black line represents the estimated conditioned mean, and the red shaded area highlights the absence of gigaton-scale deployment below 425 USD/tCO\textsubscript{2} of peak subsidies.}
    \label{fig:fig2}
\end{figure}

\section{Determinants and bottlenecks of Gt-Scale Deployment}
\label{sec:deployment}

We start by analysing the patterns of deployment of DACCS over time, the scenarios, and their determinants. The CO\textsubscript{2} removed through DACCS grows over time under both scenarios, but with distinct dynamics and only a small likelihood of reaching large-scale deployment by mid-century (Fig.~\ref{fig:fig2}, panel A). In the less ambitious NDC scenario, median CO\textsubscript{2} removals remain modest, rising from 0.04 GtCO\textsubscript{2} in 2040 to just 0.06 GtCO\textsubscript{2} by 2050. Notably, removals in 21\% of the simulations remain below 1 MtCO\textsubscript{2}, signalling a significant probability of near-zero deployment. Similarly, the likelihood of achieving gigaton-scale by mid-century is only 4\%. Under the more stringent LTS policies, median removals grow more rapidly, reaching 0.10 GtCO\textsubscript{2} in 2040 and 0.39 GtCO\textsubscript{2} by 2050. Despite these differences and longer tails, the estimated probability of getting to gigaton-scale by 2050 remains low (6\%). These findings emphasise the importance of the ambition of climate policy and the challenges of scaling up DACCS. At the same time, a long tail of high-removal outcomes exists in both scenarios, indicating that large-scale deployment remains a plausible, albeit unlikely, future.

Across both scenarios, we identify three key factors driving DACCS deployment in 2050 (Figure~\ref{fig:fig2}, panels B and C): the maximum deployment growth rate (\textit{Maximum Growth Rate}), and the level and timing of peak subsidies (\textit{Peak (Subs.)} and \textit{Timing (Subs.)}, respectively). Among these, subsidies play a pivotal role in shaping deployment trajectories under the less ambitious NDC scenario (Figure~\ref{fig:fig2}, panel B). In this context, they are needed to compensate for otherwise insufficient policy incentives. Subsidies remain influential even under the more stringent LTS scenario (Figure~\ref{fig:fig2}, panel C), although their relative importance declines as climate targets become more ambitious. This decline indicates that while robust emissions reduction policies can lessen the need for subsidies, well-designed financial support may still be relevant to accelerate learning-by-doing and unlock gigaton-scale deployment. 

In particular, the level of peak subsidies has a non-linear effect on the net removed CO\textsubscript{2} emissions, acting as a switch. No gigaton-scale deployment of DACCS occurs with subsidies below 425 USD/tCO\textsubscript{2} of peak value in both scenarios (Figures \ref{fig:fig2}, panels D and E). Above this threshold, the net removed CO\textsubscript{2} emissions distribution reaches a plateau in the LTS scenario: further increasing the peak subsidy does not lead to a proportionally larger deployment nor significant changes in the marginal distribution. This intuition is confirmed by the local separations, which measure the importance of an input when fixed at a particular value (Supplementary Figure 1-2). We also performed the sensitivity analysis by filtering only peak subsidies below 1000 USD/tCO\textsubscript{2} to check the robustness of the findings, yielding similar results (Supplementary Figure 3). We provide the partial dependency plots for the two most important inputs (\textit{Maximum Growth Rate} and \textit{Timing (Subs.)}) in Supplementary Figure 4. 

Most technological characteristics, such as capital and operational costs or energy requirements, exert a limited and often insignificant influence on DACCS deployment outcomes in both scenarios. This reflects the dominant role of subsidies and market penetration constraints in shaping deployment trajectories, in agreement with previous literature \citep{edwardsModelingDirectAir2024, realmonteIntermodelAssessmentRole2019}. A partial exception is the capacity factor, which shows a possible influence under the NDC scenario (Figure~\ref{fig:fig2}, panel B). Our results do not suggest that technological features are irrelevant, but rather that their impact is comparatively small to those of the policy environment. Taken together, these findings emphasize that the primary barriers to scaling DACCS lie not in refining technical specifications but in securing a supportive framework with stringent climate policies, targeted subsidies, and fast diffusion.

\section{Economic Implications}
\label{sec:econ}

\begin{figure}[H]
    \centering
    \includegraphics[width=1\linewidth]{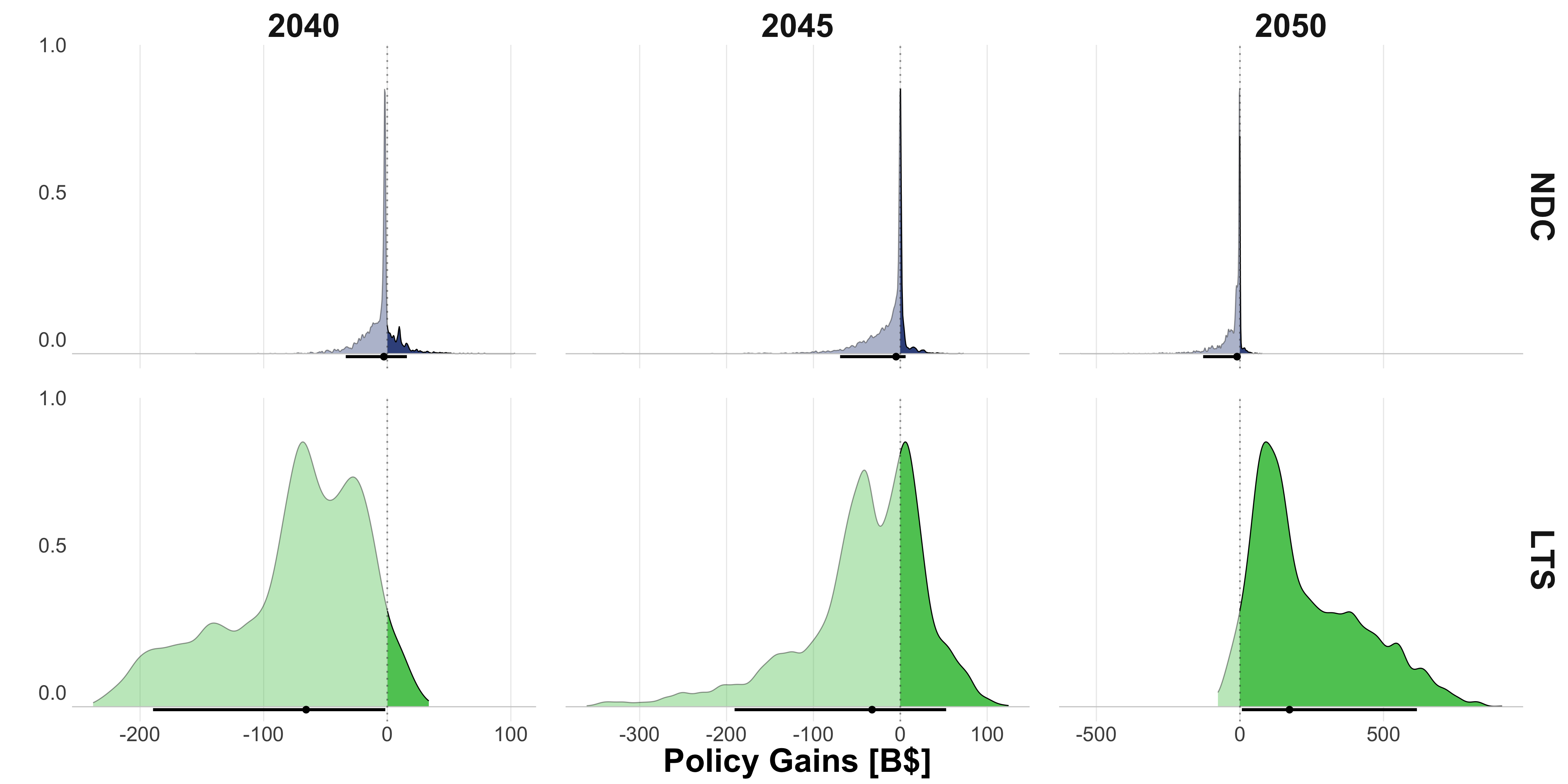}
    \caption{Distribution of policy gains measured as variation of GDP from baseline in 2040, 2045 and 2050, across scenarios. Each plot represents the probability density. Transparency identifies the zero threshold. The black bars below the plots represent medians and 5th-95th quantile ranges.}
    \label{fig:fig3}
\end{figure}

From an economic standpoint, deploying DACCS involves significant upfront costs and requires sustained support, particularly before net benefits materialize. Under the NDC scenario, global GDP policy gains are characterized by a high density close to zero and a strongly skewed distribution, with long tails (Figure~\ref{fig:fig3}). Median GDP gains range from –3 billion USD in 2040 to –10 billion USD in 2050. In contrast, median gains increase significantly under the more ambitious LTS scenario, from –66 billion USD in 2040 to 173 billion USD by 2050. Losses are even greater than in the NDC scenario in the short- to mid-term, reflecting higher levels of DACCS deployment driven by more substantial climate commitments. Still, by the mid-century, median economic benefits would be considerable (173 billion USD). In both scenarios, the likelihood of positive outcomes is non-negligible but significantly driven by the climate ambition. In the NDC scenario, the chances of positive economic benefits range from 21\% in 2040 to 6\% in 2050. In the LTS scenario, they grow from 5\% in 2040 to 95\% in 2050, meaning that by mid century it is highly likely that the public support to DACCS has been payed back. These different trends indicate, in line with the previous section, that the economic rational for supporting DACCS exist only in the presence of farsighted and ambitious climate goals; under weak policy commitments, they at best lead to transitory expansionary benefits.

Societal welfare is, however, driven by consumption more than GDP. Using this metric, we find short- and mid-term societal losses across all feasible runs (Supplementary Figure 5), compensated only in the far distant future and only in the LTS scenario. In some trajectories, net benefits exceed 4000 billion USD, highlighting the long-payback nature of DACCS (Supplementary Figure 6).

Overall, these results show that DACCS deployment may increase the burden of mitigation in the short- to mid-term. Moreover, the divergence between GDP- and consumption-based gains highlights a key feature of DACCS in the model: its deployment acts as a stimulus, boosting overall economic activity. However, these gains do not automatically translate into direct welfare benefits. The transition from stimulus for the economy to tangible welfare improvements appears only after mid-century, and only if DACCS is really needed to achieve the emission cuts compatible with net-zero.

\begin{figure}[H]
    \centering
    \includegraphics[width=1\linewidth]{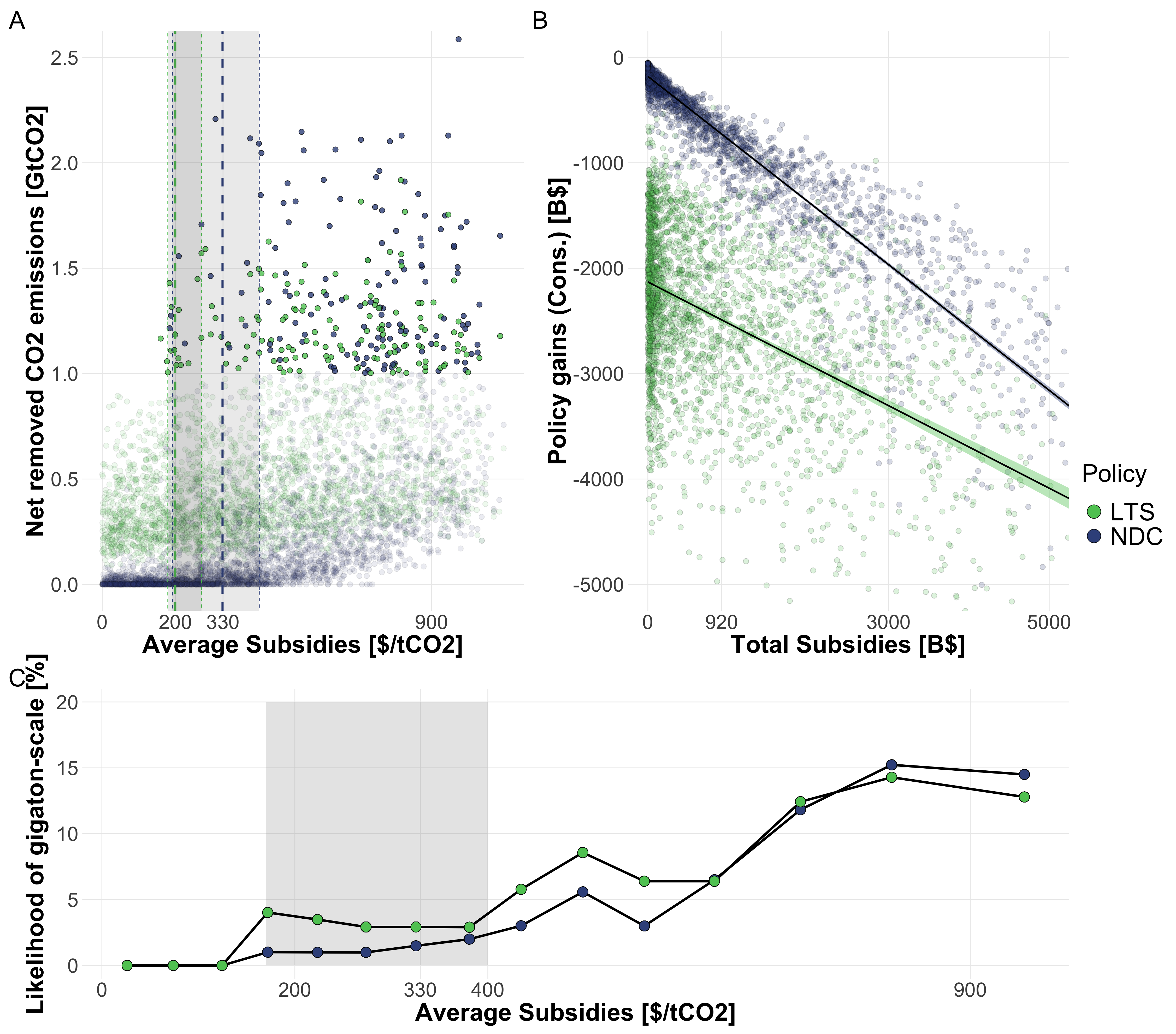}
    \caption{A. Relationship between average DACCS subsidies between 2025 and 2050 and installed capacity in 2050. The colored vertical lines represent the 5th percentile of the distribution of the points above 1 GtCO\textsubscript{2}, and the shaded area is the 95\% bootstrap confidence interval. The transparency identifies the threshold of 1 GtCO\textsubscript{2}. B. Relationship between net present value (3\% discount rate) of the DACCS subsidies distributed between 2025 and 2050 and net present value (3\% discount rate) of the consumption-based policy gains between 2025 and 2050. The black line is the smoothing performed using a linear model. Transparency identifies the zero threshold.}
    \label{fig:fig4}
\end{figure}

How much policy support is needed is key to understanding the financial desirability of a DACCS programme at scale. The two scenarios show similar minimum thresholds of subsidies to deploy gigaton-scale DACCS, around 330 USD/tCO\textsubscript{2} (95\% CI: [191, 430] USD/tCO\textsubscript{2}) in NDC scenario and 200 USD/tCO\textsubscript{2} (95\% CI: [179, 271] USD/tCO\textsubscript{2}) in LTS (Figure~\ref{fig:fig4}, panel A). Peak subsidy levels requirements are even higher. However, we find a difference between scenarios in the overall level of support required to achieve gigaton-scale deployment (Supplementary Figure 7). In the NDC scenario, scaling DACCS demands approximately 3000 billion USD in subsidies cumulatively over the next 25 years (95\% CI: [1630, 3620] billion USD). This is in the same order as the current public support for renewable power, estimated at 168 billion USD/yr \citep{laan2024public}. In contrast, under the LTS scenario, the required subsidy volume is lower, but still large at around 930 billion USD (95\% CI: [560, 1200] billion USD). The smaller amount reflects the presence of pre-existing carbon taxes and climate targets that curb the need for additional support, despite the higher volumes deployed. This level of support could be financed through the revenues generated by the carbon tax, absorbing between 1.5\% (NDC, 5th-95th quantile range: $[0, 27]$ \%) and 14\% (LTS, 5th-95th quantile range: $[0, 113]$ \%) of the fiscal space available from the climate policy, which should be earmarked for several activities, from technological to societal support. It is also worth reminding that global fossil fuel subsidies amounted to 7000 billion USD from 2010 to 2023, with a yearly average of 521 billion USD \citep{ieasubsdata, owid-how-much-subsidies-fossil-fuels}. 

This level of public support comes at societal costs. The relationship between the net present value of the subsidies and consumption losses is significantly negative (Figure~\ref{fig:fig3}, panel B), indicating that increased support to DACCS reduces overall welfare (without considering the economic benefits of lowering carbon concentrations). Notably, the regression intercept is significantly negative at –2130 billion USD, implying that, even in the absence of subsidies, the introduction of DACCS is associated with welfare losses by mid-century.

The success probability of gigaton-scale deployment positively responds to subsidies, but quite slowly and non-linearly (Figure~\ref{fig:fig3}, panel C). Exceeding the tax credit support for DACCS of the Inflation Reduction Act (180 USD/tCO\textsubscript{2}) would bring limited benefits unless governments are willing to go much higher (e.g. above 400 USD/tCO\textsubscript{2}). Still, even in those cases, chances of large-scale deployment would remain very small.

\section{Discussion}
\label{sec:concl}

We evaluate the role of direct air carbon capture and storage (DACCS) under various uncertainties, providing insights into the conditions necessary for DACCS to scale and the economic implications of its deployment. We draw on an up-to-date understanding of the uncertain inputs and advanced sensitivity analysis techniques, analysing 3000 scenarios of different input vectors through a climate-energy-economy model. 

The analysis reveals that gigaton-scale deployment is rare but possible, primarily mediated by the climate policy environment. Under weak climate goals, diffusion barriers and subsidies are critical in enabling deployment. However, as mitigation ambition increases, market penetration barriers emerge as the dominant constraint, particularly the speed at which DACCS technologies can scale. On the economic side, our results indicate that positive economic returns are achievable around mid-century following a massive public support programme, but only within ambitious climate mitigation frameworks. 

These results suggest that DACCS uptake in less ambitious policy environments is economically unjustified. In a serious policy environment, DACCS deployment may increase the mitigation burden in the mid-term in favour of long-term gains, but with significant uncertainties. Thus, promoting DACCS with public schemes sufficient for large-scale deployment must be evaluated carefully, at least on current technologies with limited maximal growth rates. Our results suggest that, if we aim for gigaton-scale, subsidies should be high for a sustained period of time (on average 200 USD/tCO\textsubscript{2}) with a high peak (at least 425 USD/tCO\textsubscript{2}). Such a programme would be financially significant, and should be well designed: subsidies exceeding the level of support contained in the US Inflation Reduction Act would bring limited additional chances of success, besides fueling economic inequality \citep{andreoniInequalityRepercussionsFinancing2024}. Besides financial incentives, to deploy DACCS at scale, policymakers will have to address diffusion barriers, such as markets' readiness to absorb and deploy carbon removal technologies, rather than focusing on decreasing the cost of the technology. While costs are more studied, these diffusion barriers are poorly understood, highly region-specific, and challenging to quantify. As a complement/alternative, new DACCS technologies with different inherent technology characteristics and consequently potentially lower diffusion barriers should be considered/supported \citep{malhotra2020accelerating}.

From a methodological viewpoint, our study highlights that uncertainty quantification and global sensitivity analysis are essential to climate change research. We have demonstrated how recent advances in global sensitivity analysis methods allow fast and robust sensitivity quantification, with indices that are simple to interpret and estimate and rely on the well-studied theoretical framework of Optimal Transport. Our results indicate that parametric uncertainty plays a significant role in shaping mitigation pathways. This and the increased computational power indicate that parametric uncertainty should be systematically explored by models and incorporated adequately into international climate science assessments such as the Intergovernmental Panel on Climate Change's Assessment Reports \citep{IPCC_AR6_WGIII_2023}.

Several limitations should be noted. First, while the implemented market growth model is standard practice in the IAM literature, it is only a high-level representation of a complex process. Two crucial, low-level factors of such a model are frictions (bureaucratic delays, environmental assessments, time to build infrastructures, and so on) and political acceptability or social resistance to DACCS deployment, which may significantly influence feasibility. Second, we do not explore structural model uncertainty, and future work should assess the robustness of these insights across alternative modelling frameworks. Moreover, the interplay between DACCS uncertainties and other permanent CDR technologies, such as BECCS and enhanced weathering, has not been explored, and it is a valuable direction for future research.

\backmatter

\section{Methods}
\label{sec:methods}

\subsection{WITCH Model}
\label{subsec:witch}

WITCH \citep{bosettiWorldInducedTechnical2006, emmerling2016witch, witchmodel} is a detailed-process Integrated Assessment Model that combines the economy, climate, and energy systems in a unified framework. In WITCH, the world is partitioned into 17 regions, denoted with $\mathcal N$. An intertemporal optimal growth model represents the economy of each region. The representation of the energy sector is hard-linked with the rest of the economy so that energy investments are chosen optimally, together with the other macroeconomic variables. The temporal horizon spans from 2020 to 2150, with 5-year timesteps.

WITCH has two relevant features to analyse emerging technologies: endogenous learning and accounting for the cost of capital. Each technology is represented by technological characteristics that can change following learning-by-doing and/or learning-by-researching frameworks. This feature is desirable when modelling novel technologies like DACCS. Moreover, WITCH has been recently expanded to account for the cost of capital \citep{calcaterra2024reducing}. This feature enables the representation of how access to affordable finances varies across technologies, time, and regions.

\subsection{Model Equations}
\label{subsec:dacmodel}

We integrate a portfolio of three direct air carbon capture and storage (DACCS) technologies into the WITCH model: liquid solvent (LS), solid sorbent (SS), and calcium oxide ambient weathering (CaO). These technologies are indexed by $d \in \{LS, SS, CaO\}$. In terms of energy inputs, SS and CaO are assumed to be fully electrified, while LS requires both electricity and natural gas.

Investments in DACCS technology $d$, at time $t$ in region $n$, are denoted as $I(d,t,n)$. Installed capacity $K(d,t,n)$ evolves according to:
\begin{equation}
\label{eq:kinv}
    K(d,t+1,n) = (1-\delta_d)K(d,t,n) + \frac{I(d,t,n)}{C_{\mathrm{cap}}(d,t,n)\omega_{\mathrm{adj}}(d,t,n)},
\end{equation}
where $\delta_d$ is the depreciation rate derived from the lifetime $LT_d$, $C_{\mathrm{cap}}(d,t,n)$ denotes the capital expenditure per unit of capacity, and $\omega_{\mathrm{adj}}(d,t,n)$ is the weighted adjusted cost of capital. The parameter $\omega_{\mathrm{adj}}(d,t,n)$ is calculated by removing the endogenous interest rates ($IR(t,n)$) from the weighted average cost of capital (WACC) denoted by $\omega(d,t,n)$ \citep{calcaterra2024reducing}:
\begin{equation}
\label{eq:wacc}
    \omega_{\mathrm{adj}}(d,t,n) = \frac{\sum_{\tau = 0}^{LT_d} (1 + IR(t + \tau,n))^{-\tau}}{\sum_{\tau = 0}^{LT_d} (1 + \omega(d,t + \tau,n))^{-\tau}}.
\end{equation}
To capture market penetration constraints, we impose a logistic growth bound on capacity expansion \citep{edwardsModelingDirectAir2024}:
\begin{equation}
\label{eq:growth_constr}
    \Delta K(d,t+1,n) \leq k K(d,t,n) \left(1 - \frac{K(d,t,n)}{L_n} \right) + K_0,
\end{equation}
with $\Delta K(d,t+1,n) = K(d,t+1,n) - K(d,t,n)$, $k$ as the maximum growth rate, $L_n$ the regional saturation level, and $K_0$ a constant enabling initial deployment in regions with no prior capacity.

Operational expenditures are captured as:
\begin{equation}
\label{eq:costemi}
    C(t,n) = \sum_d C_{\mathrm{op}}(d,t,n) K(d,t,n) + C_{\mathrm{stor}}(t,n),
\end{equation}
where $C_{\mathrm{op}}(d,t,n)$ includes costs related to labor, maintenance, and consumables, and $C_{\mathrm{stor}}(t,n)$ is the endogenous cost of storage.

Capital and operating costs, and the weighted adjusted cost of capital decline over time through endogenous learning, represented by:
\begin{equation}
\label{eq:lbd}
    C_{\mathrm{cap}}(d,t,n) = C_{\mathrm{cap}0}(d,n) \left(\frac{1}{C_{\mathrm{np}}} \sum_{\tau=1}^{t-1} \frac{I(d,\tau,n)}{C_{\mathrm{cap}}(d,\tau,n)} \right)^{b_{\mathrm{cap},d}},
\end{equation}
where $C_{\mathrm{cap}0}(d,n)$ is the initial capital cost, $C_{\mathrm{np}}$ is the nameplate capacity used to calibrate the learning parameters, and $b_{\mathrm{cap}, d}$ is the technology-specific experience rate exponent. The same equation applies to $C_{\mathrm{op}}$ and $\omega$ as well with specific initial values ($C_{\mathrm{op}0}$ and $\omega_{0}$) and learning parameters ($b_{\mathrm{op},d}$ and $b_{\mathrm{fin}}$).

Subsidies are implemented in the model as carbon taxes on top of the baseline scenario assumptions. Their functional form is:
\begin{equation} 
    \label{eq:subsidies}
        s(t) = \begin{cases}
            S\frac{t - 2025}{T - 2025},& \text{if } t\leq T\\
            S e^{-\alpha (t - T)}& \text{otherwise}
        \end{cases}\text{.}
\end{equation}
The function is parametrized by the peak $S$, the timing of the peak $T$, and the phase-out rate $\alpha$. We also constraint the maximum subsidies distributable to be at most a fraction $y_{\mathrm{frac}}$ of the regional GDP. 

\subsection{Input Calibration}
\label{subsec:incal}

The selection of the input distribution is a delicate task extensively discussed in the risk-assessment literature \citep{apostolakis1990concept, gao2016robust, chiani2025global}. We follow a standard approach: we use up-to-date distributions found in the literature when possible. Otherwise, we assign a plausible distribution, trying to explore the input space. Supplementary Table 1 contains a summary of the input distributions.

The technological characteristics are calibrated using a probabilistic DACCS cost model \citep{sievertConsideringTechnologyCharacteristics2024}. This input dimension includes: the initial capital and operational and maintenance expenditures ($C_{\mathrm{cap}0}$ and $C_{\mathrm{op}0}$), corresponding learning rates ($b_{\mathrm{cap},d}$ and $b_{\mathrm{op},d}$), capacity factors ($f_d$), energy requirements ($\zeta$ and $\eta$), and lifetimes ($LT_d$). 

Financing costs lack probabilistic characterizations in the literature, and no data exist on the weighted average cost of capital (WACC) for DACCS. We therefore assign to the WACC for Europe a truncated Gaussian distribution centred on values reported in related studies \citep{fasihiTechnoeconomicAssessmentCO22019, sievertConsideringTechnologyCharacteristics2024}, allowing for a broad range of possible WACCs. For the other 16 regions, we sample from other technologies' WACC data and apply the same relative regional differences. Finally, we assume a uniform distribution centred on the default value for the learning-by-financing input $b_{\mathrm{fin}}$.

The market penetration constraint follows the logistic growth constraint (Equation \eqref{eq:growth_constr}). For the maximum market growth rate ($k$), we base the distribution on technological analogue data \citep{edwardsModelingDirectAir2024}, filter out values below 0.09 to reduce infeasibilities, and generate a continuous distribution using a kernel density estimator \citep{duong2007ks}. We assume a uniform distribution for the initial capacity $K_0$, reflecting the absence of prior knowledge and aligning with GSA practice in IAMs \citep{anderson2014uncertainty}. The regional saturation level $L_n$ is derived in three steps. First, we identify areas with a mean annual temperature above –15°C \citep{sendi2022geospatial} using historical data \citep{copernicusAtlas}. Second, we estimate available land by country suitable for DACCS siting using FAO land-use data \citep{fao2025landusestats, faodata}. Third, we combine these with DACCS land-use factors \citep{motlaghzadehKeyUncertaintiesGlobal2023} to determine maximum installable capacity per region. These values serve as regional capacity shares, while the global saturation level is drawn from a uniform distribution centred on the maximum value in the AR6 database \citep{edwardsModelingDirectAir2024}. While simplified, this approach offers a first-order estimate of a critical constraint in DACCS deployment modelling. Supplementary Table 2 displays the range of the global and regional saturation levels.

Subsidies are modelled as a simple parametric function governed by four intuitive variables: timing ($T$), peak value ($S$), phase-out rate ($\alpha$), and maximum subsidies distributable ($y_{\mathrm{frac}}$). Since these inputs are normative, we design distributions to explore the space thoroughly rather than be policy-realistic. We assume a uniform distribution for the $T$, $S$, and $y_{\mathrm{frac}}$, while we assign to $\alpha$ a gamma distribution to address the possibility of fast decays in policy support. We assume that only regions with reasonably ambitious LTS policies subsidize DACCS, defined as an emissions reduction of at least 50\% by the end of the century. By using this threshold, we filter out five of the seventeen regions: \texttt{laca} (Latin America and Caribbean), \texttt{mena} (Middle East and North Africa), \texttt{mexico} (Mexico), \texttt{sasia} (South Asia, excluding India), \texttt{ssa} (Sub-Saharan Africa), \texttt{te} (Non-EU Eastern European countries). 

\subsection{Quantities of Interest}
\label{subsec:qoi}

We consider three quantities of interest: the yearly net removed CO\textsubscript{2} emissions, the policy gains, and the total subsidies. The yearly net removed CO\textsubscript{2} emissions are defined as:
\begin{equation}
    \label{eq:netemi}
    E(t) = \sum_{n \in \mathcal N} \sum_d f_d K(d,t,n) \quad \text{for } t \in \{2040, 2045, 2050\},
\end{equation}
where $K$ is the installed capacity (Equation~\eqref{eq:kinv}), and $f_d$ is the capacity factor. In the uncertainty quantification, we separately analyse the three quantities of interest. In the GSA step, we consider the multivariate output $(E(2040), E(2045), E(2050))$.

The yearly policy gains for a scenario $s \in \{NDC, LTS\}$ are defined as:
\begin{equation}
    \label{eq:ypolgains}
    G(t) = \sum_{n \in \mathcal N} \left[ Q(t,n) - Q_s (t,n) \right] \quad \text{for } t \in \{2025, 2030, \dots, 2050\},
\end{equation}
where $Q$ is either the consumption or the GDP in the specific Monte Carlo run, and $Q_s$ is the consumption or GDP in a run for scenario $s$ with small DACCS deployment. When aggregating in time, we consider the net present value as:
\begin{equation}
\label{eq:polgains}
    G = \sum_{t=2025}^{2050} G(t) \left( \sum_{\tau = t}^{t + 4} \frac{1}{(1+\rho)^{\tau - 2025}} \right),
\end{equation}
We assume a discount rate of 3\% per year and account for the 5-year timesteps through the sum indexed by $\tau$.

The total subsidies are the actual subsidies handed out, in the form:
\begin{equation}
    \label{eq:totsubs}
    TS = \sum_{t=2025}^{2050} \sum_{n \in \mathcal N} \sum_d f_d K(d,t,n) s(t) \left( \sum_{\tau = t}^{t + 4} \frac{1}{(1+\rho)^{\tau - 2025}} \right).
\end{equation}

\subsection{Optimal Transport-based Sensitivity Indices}
\label{subsec:gsaot}

Here, we present a succinct overview of the subject. We refer to more detailed works for an extended discussion on Optimal Transport \citep{villani2009, peyre2019computational}, the Optimal Transport-based sensitivity indices \citep{wiesel2022, borgonovo2024global}, and the computational aspects of indices estimation \cite{peyre2019computational, chiani2025gsaot}. 

Let us consider two probability distributions, $\mathbb{P}$ and $\mathbb{P}'$, defined over the same space $\mathcal{Y} \subseteq \mathbb R^m$. Let us also define a ground cost $k\colon \mathcal{Y} \times \mathcal{Y} \longrightarrow [0, +\infty]$ over the space $\mathcal{Y}$. The optimal transport (OT) problem can be expressed as the problem of finding the least-cost coupling between $\mathbb{P}$ and $\mathbb{P}'$ given the ground cost function. The couplings, also called transport plans, are joint probability measures on $\mathcal{Y} \times \mathcal{Y}$, such that their marginals are $\mathbb{P}$ and $\mathbb{P}'$. 

When the ground cost is the squared Euclidean distance, the OT problem defines a distance over the space of probability measures, called the Wasserstein distance, that can be expressed as:
\begin{equation}
\begin{aligned}
\label{eq:wdecomp}
    K(\mathbb{P}, \mathbb{P}') =& \operatorname{WB}(\mathbb{P}, \mathbb{P}') + \Gamma(\mathbb{P}, \mathbb{P}') \\=& \lVert m - m' \rVert^2_2 + \operatorname{Tr} \biggl( \Sigma + \Sigma' - 2 \Bigl( \Sigma'^{\frac{1}{2}} \Sigma \Sigma'^{\frac{1}{2}} \Bigr)^{\frac{1}{2}} \biggl) + \Gamma(\mathbb{P}, \mathbb{P}'),
\end{aligned}
\end{equation}
where $m$, $m'$ and $\Sigma$, $\Sigma'$ are the means and covariance matrices of $\mathbb{P}$ and $\mathbb{P}'$ respectively, and $\operatorname{Tr}$ is the trace operator, summing all diagonal entries of a square matrix. We call $K(\mathbb{P}, \mathbb{P}')$ the OT cost. The sum of the first two terms on the second line of Equation~\eqref{eq:wdecomp} defines the Wasserstein–Bures semi-metric, denoted by $\operatorname{WB}(\cdot,\cdot)$. The first term captures the cost of aligning the mean values of the distributions $\mathbb{P}$ and $\mathbb{P}'$, corresponding to their first-order moments. The second term represents the optimal cost of matching their variance-covariance matrices, i.e., the second-order moments. The third term, $\Gamma(\cdot,\cdot)$, accounts for the differences in higher-order moments. We refer to it as the residual term, which generally lacks a closed-form expression. As a result, the full Wasserstein distance between two arbitrary distributions cannot be derived analytically.

Let us consider a model $f$ with inputs $\mathbf X = (X_1, \dots, X_n)$ and (possibly multivariate) output $\mathbf Y \in \mathcal{Y}$. Let us further define the unconditioned output distribution as $\mathbb{P}_{\mathbf{Y}}$, and the distribution of the output fixed one input $X_i$ as $\mathbb{P}_{\mathbf{Y} | X_i}$. The OT cost $K(\mathbb{P}, \mathbb{P}')$ can be used to define a measure of statistical association between the output $\mathbf Y$ and the input $X_i$ as:
\begin{equation} \label{eq:smot}
  \xi^K (\mathbf{Y}, X_i) = \mathbb E_{X_i}[K(\mathbb{P}_{\mathbf{Y}}, \mathbb{P}_{\mathbf{Y} | X_i})]\text{.}
\end{equation}
In this case, we call the function $\gamma(X_i) = K(\mathbb{P}_{\mathbf{Y}}, \mathbb{P}_{\mathbf{Y} | X_i})$ local separation measure. We can derive an upper bound for $\xi^K (\mathbf{Y}, X_i)$:
\begin{equation} \label{eq:upbound}
    \xi^K (\mathbf{Y}, X_i) \leq 2\mathbb{V}[\mathbf{Y}],
\end{equation}
where $\mathbb{V}[\mathbf{Y}]$ is the sum of the diagonal elements of the variance-covariance matrix of the output(s) $\mathbf{Y}$. Since $\mathbb{V}[\mathbf{Y}] > 0$, we can further define the Optimal Transport-based sensitivity index (OT-based index) as:
\begin{equation}
\label{eq:otindex}
    \iota^{K} (\mathbf{Y}, X_i) = \frac{\xi^{K} (\mathbf{Y}, X_i)}{2\mathbb{V}[\mathbf{Y}]}.
\end{equation}
This index has several desirable properties described in \cite{moriFourSimpleAxioms2019a}. First, $\iota^K (\mathbf{Y}, X_i) \geq 0$, and $\iota^K (\mathbf{Y}, X_i) = 0$ if and only if $\mathbf{Y}$ and $X_i$ are independent. This is called the zero-independence property. Second, $\iota^K (\mathbf{Y}, X_i) \leq 1$ and $\iota^K (\mathbf{Y}, X_i) = 1$ if and only if there exists a measurable function $\mathbf{g}$ such that $\mathbf{Y}=\mathbf{g}(X_i)$. This second property is called max-functionality. Together, these two properties enable the indices to give synthetic but comprehensive information on the statistical association between two random variables. Moreover, the decomposition in Equation~\eqref{eq:wdecomp} enables the decomposition of the OT-based indices as $\iota^K (\mathbf{Y}, X_i) = \iota^{V}(\mathbf{Y},X_{i})+\iota ^{\Sigma}(\mathbf{Y},X_{i})+\iota ^{\Gamma}(\mathbf{Y},X_{i})$. Here, $\iota^{V}(\mathbf{Y},X_{i})$ estimates the importance of $X_{i}$ on the mean of $\mathbf{Y}$, $\iota ^{\Sigma}(\mathbf{Y},X_{i})$ quantifies the importance of $X_{i}$ on the variance-covariance matrix of $\mathbf{Y}$, and $\iota ^{\Gamma}(\mathbf{Y},X_{i})$ identifies the higher order effects.

Even though the indices possess the zero-independence property, the numerical estimation may induce some non-zero index values even in the case of independence. The answer to the question of whether these inputs are influential depends on understanding whether the non-null estimates result from numerical noise. We address this problem by introducing an auxiliary random variable $X_{dummy}$, independent of the output $\mathbf Y$ by construction, and computing its OT-based index $\iota^K (\mathbf{Y}, X_{dummy})$. Given the zero-independence property, a non-zero estimate of $\iota^K(\mathbf Y, X_{dummy})$ is due to numerical noise. Thus, we can use this value as an irrelevance threshold.

\subsection{Design of experiment}
\label{subsec:doe}

We generate 3,000 input vectors drawing from a joint distribution constructed from the marginal distributions shown in Supplementary Table 1. The data-driven inputs are sampled using the probabilistic model accompanying the related work \citep{sievertConsideringTechnologyCharacteristics2024}. The \textit{Maximum Growth Rate} ($k$) is sampled using a kernel density estimator implemented in the \texttt{ks} R package \citep{ks}. The remaining parameters are sampled using Latin Hypercube Sampling (LHS). Specifically, we generate 30 LHS designs with 100 points each using the \texttt{FME} R package \citep{soetaert2010inverse}, transforming the uniformly distributed samples into the target distributions via the inverse transform method.

To leverage high-performance computing capabilities, the full experimental design is partitioned into 100 equally sized clusters to enable parallel execution. Before clustering, all inputs are transformed into quantile space to ensure uniform scaling. Cluster centroids are generated using a Sobol’ sequence \citep{joe2008constructing} to ensure good space-filling properties. The clustering is formulated as a balanced optimal transport problem, assigning each design point to its nearest centroid while preserving uniform cluster sizes. The optimal transport problem is solved using the simplex method implemented by the \texttt{transport} R package \citep{transport}. Let $\{x_i\}_{i=1}^{3000}, \{c_j\}_{j=1}^{100} \subset [0,1]^{36}$ be two sets of points. Defined the cost of assigning a point $x_i$ to a centroid $c_j$ as $k_{ij} = \|x_i - c_j\|^2$, the mathematical formulation of the problem is:
\begin{equation}
\begin{aligned}
    \min_{\pi_{ij}} \quad & \sum_{i=1}^{3000} \sum_{j=1}^{100} k_{ij} \pi_{ij} \\
    \text{s.t.} \quad
    & \sum_{j=1}^{100} \pi_{ij} = \frac{1}{3000} \quad \forall i = 1, \dots, 3000 \\
    & \sum_{i=1}^{3000} \pi_{ij} = \frac{1}{100} \quad \forall j = 1, \dots, 100 \\
    & 0 \leq \pi_{ij} \leq 1 \quad \forall i, j
\end{aligned}
\end{equation}
Given the non-linear nature of the WITCH model, convergence speed is improved by supplying informed initial values. For each scenario, we first run the model using the mean values of all inputs. Then, for each cluster, a greedy nearest-neighbour algorithm orders the 30 points starting from the mean point, minimizing the Euclidean square distance in quantile space. This ordering promotes faster convergence during model execution. The result is a set of 100 parallelizable and internally ordered clusters, each composed of 30 design points. The model is solved using the CONOPT3 solver on 3rd-generation Intel Xeon Scalable processors.
Then, we compute the sensitivity indices using the \texttt{gsaot} R package \citep{chiani2025gsaot}. The whole pipeline is implemented using the \texttt{targets} R package \citep{targets}, enabling reproducibility and interpretability.

\bmhead{Acknowledgements}
L.C. and M.T. acknowledge support from the European Union ERC Consolidator Grant under project No. 101044703 (EUNICE). P.A. and L.D. acknowledge support from the European Union's Horizon Europe programme under grant agreement No 101081521 (UPTAKE).
As part of the PRISMA project, B.S., T.S. and K.S. received funding from the Swiss State Secretariat for Education, Research and Innovation (SERI, contract no. 22.00541).

\section*{Declarations}

Replication code and data are available at https://zenodo.org/records/17458632.




\bibliography{sn-bibliography}

\end{document}


\title[\textbf{Supplementary Information} \\ The Uncertain Policy Price of Scaling Direct Air Capture]{\centering \textbf{Supplementary Information} \\ The Uncertain Policy Price of Scaling Direct Air Capture}


\author*[1,2,3]{\fnm{Leonardo} \sur{Chiani}\email{leonardo.chiani@polimi.it}} 
\author[1,2,3]{\fnm{Pietro} \sur{Andreoni}}
\author[2,3]{\fnm{Laurent} \sur{Drouet}}
\author[4]{\fnm{Tobias} \sur{Schmidt}}
\author[4,5]{\fnm{Katrin} \sur{Sievert}}
\author[4,5,6]{\fnm{Bjarne} \sur{Steffen}}
\author[1,2,3]{\fnm{Massimo} \sur{Tavoni}}
\affil*[1]{\footnotesize Department of Management Engineering, Politecnico di Milano, Italy}
\affil[2]{\footnotesize CMCC Foundation - Euro-Mediterranean Center on Climate Change, Italy}
\affil[3]{\footnotesize RFF-CMCC European Institute on Economics and the Environment, Italy}
\affil[4]{\footnotesize Energy and Technology Policy Group, ETH Zurich, Switzerland}
\affil[5]{\footnotesize Climate Finance and Policy Group, ETH Zurich, Switzerland}
\affil[6]{\footnotesize Center for Energy and Environmental Policy Research, Massachusetts Institute of Technology, USA}



\maketitle

\section{Table of Input Parameters}\label{secA1}

\begin{sidewaystable}[htp]
    \centering
    \begin{adjustbox}{width=1.1\textwidth}
    \begin{tabular}{c|c|c|c|c|c}
        \textbf{Input} & \textbf{Symbol} & \textbf{Technology} & \textbf{Distribution} & \textbf{Parameters} & \textbf{Source} \\
        \midrule
        Initial CAPEX & $C_{cap0}(d,n)$ & LS & Uniform & $[2067.01, 3510.25] \quad 2022USD / tCO2$ & \cite{sievertConsideringTechnologyCharacteristics2024} \\
        Initial CAPEX & $C_{cap0}(d,n)$ & SS & Delta & $5698.37\quad 2022USD / tCO2$ & \cite{sievertConsideringTechnologyCharacteristics2024} \\
        Initial CAPEX & $C_{cap0}(d,n)$ & CaO & Uniform & $[11212.50, 13532.06] \quad 2022USD / GtCO2$ & \cite{sievertConsideringTechnologyCharacteristics2024} \\
        Initial OPEX & $C_{op0}(d,n)$ & LS & Data-driven & $[182.6,   221.3,   243.1,   247.9,   271.8,   353.2] \quad 2022USD/tCO2$ & \cite{sievertConsideringTechnologyCharacteristics2024}, they are $f_dVOM+FOM$ \\
        Initial OPEX & $C_{op0}(d,n)$ & SS & Data-driven & $[468.4,   525.8,   547.5,   547.5,   569.3,   625.5] \quad 2022USD/tCO2$ & \cite{sievertConsideringTechnologyCharacteristics2024}, they are $f_dVOM+FOM$ \\
        Initial OPEX & $C_{op0}(d,n)$ & CaO & Data-driven & $[630.8,   722.3,   762.5,   765.9,   808.1,   935.1] \quad 2022USD/tCO2$ & \cite{sievertConsideringTechnologyCharacteristics2024}, they are $f_dVOM+FOM$ \\
        Learning (CAPEX) & $b_{cap,d}$ & LS & Data-driven & $[-0.011,  0.076,  0.097,  0.097,  0.118,  0.197]$ & \cite{sievertConsideringTechnologyCharacteristics2024} \\
        Learning (CAPEX) & $b_{cap,d}$ & SS & Data-driven & $[-0.001,  0.111,  0.125,  0.122,  0.136,  0.171]$ & \cite{sievertConsideringTechnologyCharacteristics2024} \\
        Learning (CAPEX) & $b_{cap,d}$ & CaO & Data-driven & $[-0.008,  0.119,  0.143,  0.138,  0.162,  0.211]$ & \cite{sievertConsideringTechnologyCharacteristics2024} \\
        Learning (OPEX) & $b_{op,d}$ & LS & Data-driven & $[-0.109,  0.048,  0.082,  0.080,  0.118,  0.230]$ & \cite{sievertConsideringTechnologyCharacteristics2024} \\
        Learning (OPEX) & $b_{op,d}$ & SS & Data-driven & $[-0.100,  0.065,  0.080,  0.077,  0.092,  0.137]$ & \cite{sievertConsideringTechnologyCharacteristics2024} \\
        Learning (OPEX) & $b_{op,d}$ & CaO & Data-driven & $[-0.140,  0.121,  0.146,  0.138,  0.166,  0.225]$ & \cite{sievertConsideringTechnologyCharacteristics2024} \\
        Minimum cost & $C_{cap,min}$ & Unique & Uniform & $[0, 100] \quad \$/tCO2$ & Our Assumption \\
        Thermal Cons. & $\zeta_d$ & LS & Truncated Normal & $(5.3, 0.541, 2.67, +\infty) \quad GJ/tCO2$ & Central value from \cite{sievertConsideringTechnologyCharacteristics2024}, the distribution is our assumption \\
        Thermal Cons. & $\zeta_d$ & SS & Truncated Normal & $(9.8, 1, 4, +\infty) \quad GJ/tCO2$ & Central value from \cite{sievertConsideringTechnologyCharacteristics2024}, the distribution is our assumption \\
        Thermal Cons. & $\zeta_d$ & CaO & Delta & $0 \quad GJ/tCO2$ & Technological characteristic \\
        Electric Cons. & $\eta_d$ & LS & Truncated Normal & $(1.32, 0.135, 0.8, +\infty) \quad GJ/tCO2$ & Central value from \cite{sievertConsideringTechnologyCharacteristics2024}, the distribution is our assumption \\
        Electric Cons. & $\eta_d$ & SS & Truncated Normal & $(0.99, 0.101, 0.4, +\infty) \quad GJ/tCO2$ & Central value from \cite{sievertConsideringTechnologyCharacteristics2024}, the distribution is our assumption \\
        Electric Cons. & $\eta_d$ & CaO & Truncated Normal & $(9, 0.918, 2.4, +\infty) \quad GJ/tCO2$ & Central value from \cite{sievertConsideringTechnologyCharacteristics2024}, the distribution is our assumption \\
        Capacity Factor & $f_d$ & LS & Uniform & $[0.5, 0.9]$ & \cite{sievertConsideringTechnologyCharacteristics2024} \\
        Capacity Factor & $f_d$ & SS & Uniform & $[0.75, 0.9]$ & \cite{sievertConsideringTechnologyCharacteristics2024} \\
        Capacity Factor & $f_d$ & CaO & Uniform & $[0.5, 0.9]$ & \cite{sievertConsideringTechnologyCharacteristics2024} \\
        Lifetime & $LT_d$ & LS & Discrete Uniform & $\{20, \dots, 25\} \quad \text{years}$ & Minimum from \cite{madhu2021understanding}, maximum from \cite{sievertConsideringTechnologyCharacteristics2024} \\
        Lifetime & $LT_d$ & SS & Discrete Uniform & $\{20, \dots, 25\} \quad \text{years}$ & Minimum from \cite{madhu2021understanding}, maximum from \cite{sievertConsideringTechnologyCharacteristics2024} \\
        Lifetime & $LT_d$ & CaO & Discrete Uniform & $\{20, \dots, 25\} \quad \text{years}$ & Minimum from \cite{madhu2021understanding}, maximum from \cite{sievertConsideringTechnologyCharacteristics2024} \\
        \midrule
        Maximum Growth Rate & $k$ & No & Data-driven & $[0.06691, 0.09154, 0.13406, 0.15689, 0.21664, 0.34038]$ & Kernel-density estimation from \cite{edwardsModelingDirectAir2024} \\
        Initial Capacity & $K_0$ & No & Uniform & $[0.8, 1.2] \quad MtCO2$ & Arbitrary \\
        Maximum Capacity & $L_n$ & No & Uniform & $[0.005, 0.03]$ & Land used and mean temperature, check Table \ref{tab:dac_kmax} for regional specifics \\
        \midrule
        Peak (Subs.) & $S$ & No & Uniform & $[0, 1800] \quad 2005USD/tCO2$ & Normative \\
        Timing (Subs.) & $T$ & No & Discrete Uniform & $\{2025,2030,\dots,2050\}$ & Normative \\
        Phase-Out Rate (Subs.) & $\alpha$ & No & Gamma & $(7,7)$ & Normative \\
        Max. Subs. Distributable & $y_{frac}$ & No & Uniform & $[0.005, 0.05] \text{\% of regional GDP}$ & Normative \\
        \midrule
        Cost of Capital & $\omega_d$ & LS & Truncated Normal & $(0.07, 0.03, 0, +\infty)$ & Central value from \cite{sievertConsideringTechnologyCharacteristics2024, fasihiTechnoeconomicAssessmentCO22019}, the distribution is our assumption\\
        Cost of Capital & $\omega_d$ & SS & Truncated Normal & $(0.07, 0.03, 0, +\infty)$ & Central value from \cite{sievertConsideringTechnologyCharacteristics2024, fasihiTechnoeconomicAssessmentCO22019}, the distribution is our assumption\\
        Cost of Capital & $\omega_d$ & CaO & Truncated Normal & $(0.07, 0.03, 0, +\infty)$ & Central value from \cite{sievertConsideringTechnologyCharacteristics2024, fasihiTechnoeconomicAssessmentCO22019}, the distribution is our assumption\\
        Learning (Capital) & $b_{fin}$ & No & Uniform & $[0.02, 0.08]$ & Central value from \cite{calcaterra2024reducing}, the distribution is our assumption\\
        Convergence Rate & $\omega_{conv}$ & No & Log-Normal & $(0,1)$ & Our assumption\\
        Regional Spread & $\omega_{reg}$ & No & Discrete Uniform & $\{1,\dots,10\}$ & Our assumption
    \end{tabular}
    \end{adjustbox}
    \caption{Table with currently implemented inputs and their distribution (not the most entertaining thing possible, however). The parameters column has different meanings for the different distributions: (\textit{Uniform}) the parameters are the minimum and maximum of the distribution (\textit{Delta}) the parameter is the single point (\textit{Data-driven}) the parameters are the minimum, 1st quartile, median, mean, 3rd quartile, and maximum values (\textit{Discrete Uniform}) the parameters are representative of the set (\textit{Gamma}) the parameters are the shape and rate (\textit{Normal}) the parameters are the mean and the standard deviation (\textit{Log-Normal}) the parameters are the mean and standard deviation in log-scale.}\label{tab:inputs}
\end{sidewaystable}

\begin{table}[htp]
\centering
\begin{tabular}{c|c|c|c}
 \textbf{Region} & \textbf{Central Value} & \textbf{Lower Bound} & \textbf{Upper Bound} \\ 
  \midrule
    brazil & 0.88 & 0.25 & 1.52 \\ 
    canada & 0.74 & 0.21 & 1.27 \\ 
    china & 1.14 & 0.33 & 1.95 \\ 
    europe & 1.00 & 0.28 & 1.71 \\ 
    india & 0.40 & 0.12 & 0.69 \\ 
    indonesia & 0.29 & 0.08 & 0.50 \\ 
    jpnkor & 0.10 & 0.03 & 0.18 \\ 
    laca & 2.46 & 0.70 & 4.21 \\ 
    mena & 6.51 & 1.86 & 11.16 \\ 
    mexico & 0.28 & 0.08 & 0.48 \\ 
    oceania & 2.32 & 0.66 & 3.97 \\ 
    sasia & 0.60 & 0.17 & 1.03 \\ 
    seasia & 0.52 & 0.15 & 0.89 \\ 
    southafrica & 0.07 & 0.02 & 0.12 \\ 
    ssa & 6.16 & 1.76 & 10.56 \\ 
    te & 1.74 & 0.50 & 2.98 \\ 
    usa & 1.41 & 0.40 & 2.42 \\ 
    global & 26.62 & 7.61 & 45.64 
\end{tabular}
\caption{Regional and global maximum market capacities in GtCO2.}
\label{tab:dac_kmax}
\end{table}

\section{Additional Sensitivity Results on Emissions}

\begin{table}[ht]
\centering
\resizebox{\textwidth}{!}{\begin{tabular}{lllrrr}
\textbf{Input} & \textbf{DAC Technology} & \textbf{Dimension} & \textbf{OT-based Index} & \textbf{CI (low)} & \textbf{CI (high)} \\ 
  \hline
Maximum Growth Rate & Global & Market & 0.27 & 0.24 & 0.29 \\ 
  Timing (Subs.) & Global & Political & 0.27 & 0.24 & 0.29 \\ 
  Peak (Subs.) & Global & Political & 0.20 & 0.18 & 0.22 \\ 
  Phase-Out Rate (Subs.) & Global & Political & 0.07 & 0.05 & 0.08 \\ 
  Max. Subs. Distributable & Global & Political & 0.06 & 0.04 & 0.07 \\ 
  Initial Capacity & Global & Market & 0.05 & 0.04 & 0.07 \\ 
  Capacity Factor & LS & Technical & 0.05 & 0.04 & 0.06 \\ 
  Maximum (Capacity) & Global & Market & 0.04 & 0.03 & 0.05 \\ 
  Elec. Cons. & LS & Technical & 0.04 & 0.03 & 0.05 \\ 
  Cost of Capital & LS & Finance & 0.04 & 0.03 & 0.05 \\ 
  Inital OPEX & SS & Technical & 0.04 & 0.03 & 0.05 \\ 
  Learning (OPEX) & LS & Technical & 0.04 & 0.03 & 0.05 \\ 
  Learning (OPEX) & SS & Technical & 0.04 & 0.03 & 0.05 \\ 
  Learning (CAPEX) & CaO & Technical & 0.04 & 0.03 & 0.05 \\ 
  Capacity Factor & CaO & Technical & 0.04 & 0.03 & 0.05 \\ 
  Initial CAPEX & LS & Technical & 0.04 & 0.03 & 0.05 \\ 
  Learning (Capital) & Global & Finance & 0.04 & 0.03 & 0.05 \\ 
  Learning (CAPEX) & SS & Technical & 0.04 & 0.03 & 0.05 \\ 
  Lifetime & SS & Technical & 0.04 & 0.03 & 0.04 \\ 
  Inital OPEX & LS & Technical & 0.04 & 0.03 & 0.05 \\ 
  Inital OPEX & CaO & Technical & 0.04 & 0.02 & 0.04 \\ 
  Lifetime & LS & Technical & 0.04 & 0.03 & 0.04 \\ 
  Thermal Cons. & SS & Technical & 0.04 & 0.03 & 0.04 \\ 
  Cost of Capital & SS & Finance & 0.04 & 0.02 & 0.04 \\ 
  Learning (CAPEX) & LS & Technical & 0.04 & 0.02 & 0.04 \\ 
  Capacity Factor & SS & Technical & 0.04 & 0.02 & 0.04 \\ 
  Elec. Cons. & SS & Technical & 0.03 & 0.02 & 0.04 \\ 
  Minimum Cost & Global & Technical & 0.03 & 0.02 & 0.04 \\ 
  Initial CAPEX & CaO & Technical & 0.03 & 0.02 & 0.04 \\ 
  Elec. Cons. & CaO & Technical & 0.03 & 0.02 & 0.04 \\ 
  Cost of Capital & CaO & Finance & 0.03 & 0.02 & 0.04 \\ 
  Lifetime & CaO & Technical & 0.03 & 0.03 & 0.04 \\ 
  Learning (OPEX) & CaO & Technical & 0.03 & 0.02 & 0.04 \\ 
  Thermal Cons. & LS & Technical & 0.03 & 0.02 & 0.04 
\end{tabular}}
\caption{OT-based indices for the net removed CO\textsubscript{2} emissions in NDC scenario.}
\end{table}

\begin{table}[ht]
\centering
\resizebox{\textwidth}{!}{\begin{tabular}{lllrrr}
\textbf{Input} & \textbf{DAC Technology} & \textbf{Dimension} & \textbf{OT-based Index} & \textbf{CI (low)} & \textbf{CI (high)} \\ 
  \hline
Maximum Growth Rate & Global & Market & 0.42 & 0.40 & 0.43 \\ 
  Timing (Subs.) & Global & Political & 0.13 & 0.11 & 0.14 \\ 
  Peak (Subs.) & Global & Political & 0.10 & 0.08 & 0.11 \\ 
  Phase-Out Rate (Subs.) & Global & Political & 0.06 & 0.05 & 0.07 \\ 
  Initial Capacity & Global & Market & 0.06 & 0.05 & 0.07 \\ 
  Max. Subs. Distributable & Global & Political & 0.06 & 0.04 & 0.06 \\ 
  Maximum (Capacity) & Global & Market & 0.05 & 0.04 & 0.06 \\ 
  Capacity Factor & CaO & Technical & 0.05 & 0.04 & 0.06 \\ 
  Capacity Factor & LS & Technical & 0.05 & 0.04 & 0.06 \\ 
  Learning (OPEX) & CaO & Technical & 0.05 & 0.04 & 0.06 \\ 
  Elec. Cons. & LS & Technical & 0.05 & 0.04 & 0.06 \\ 
  Inital OPEX & CaO & Technical & 0.05 & 0.04 & 0.06 \\ 
  Elec. Cons. & CaO & Technical & 0.05 & 0.04 & 0.06 \\ 
  Initial CAPEX & LS & Technical & 0.05 & 0.04 & 0.06 \\ 
  Learning (Capital) & Global & Finance & 0.05 & 0.04 & 0.06 \\ 
  Capacity Factor & SS & Technical & 0.05 & 0.04 & 0.06 \\ 
  Learning (OPEX) & SS & Technical & 0.05 & 0.04 & 0.06 \\ 
  Thermal Cons. & SS & Technical & 0.05 & 0.04 & 0.05 \\ 
  Elec. Cons. & SS & Technical & 0.05 & 0.04 & 0.06 \\ 
  Learning (OPEX) & LS & Technical & 0.05 & 0.04 & 0.05 \\ 
  Cost of Capital & LS & Finance & 0.05 & 0.04 & 0.05 \\ 
  Inital OPEX & SS & Technical & 0.05 & 0.04 & 0.05 \\ 
  Lifetime & SS & Technical & 0.05 & 0.04 & 0.05 \\ 
  Learning (CAPEX) & CaO & Technical & 0.05 & 0.04 & 0.05 \\ 
  Inital OPEX & LS & Technical & 0.05 & 0.04 & 0.05 \\ 
  Cost of Capital & CaO & Finance & 0.05 & 0.04 & 0.05 \\ 
  Lifetime & LS & Technical & 0.05 & 0.04 & 0.05 \\ 
  Learning (CAPEX) & SS & Technical & 0.05 & 0.04 & 0.05 \\ 
  Cost of Capital & SS & Finance & 0.05 & 0.04 & 0.05 \\ 
  Initial CAPEX & CaO & Technical & 0.04 & 0.04 & 0.05 \\ 
  Lifetime & CaO & Technical & 0.04 & 0.04 & 0.05 \\ 
  Learning (CAPEX) & LS & Technical & 0.04 & 0.04 & 0.05 \\ 
  Minimum Cost & Global & Technical & 0.04 & 0.04 & 0.05 \\ 
  Thermal Cons. & LS & Technical & 0.04 & 0.03 & 0.05 
\end{tabular}}
\caption{OT-based indices for the net removed CO\textsubscript{2} emissions in LTS scenario.}
\end{table}

\begin{figure}[H]
    \centering
    \includegraphics[width=1\linewidth]{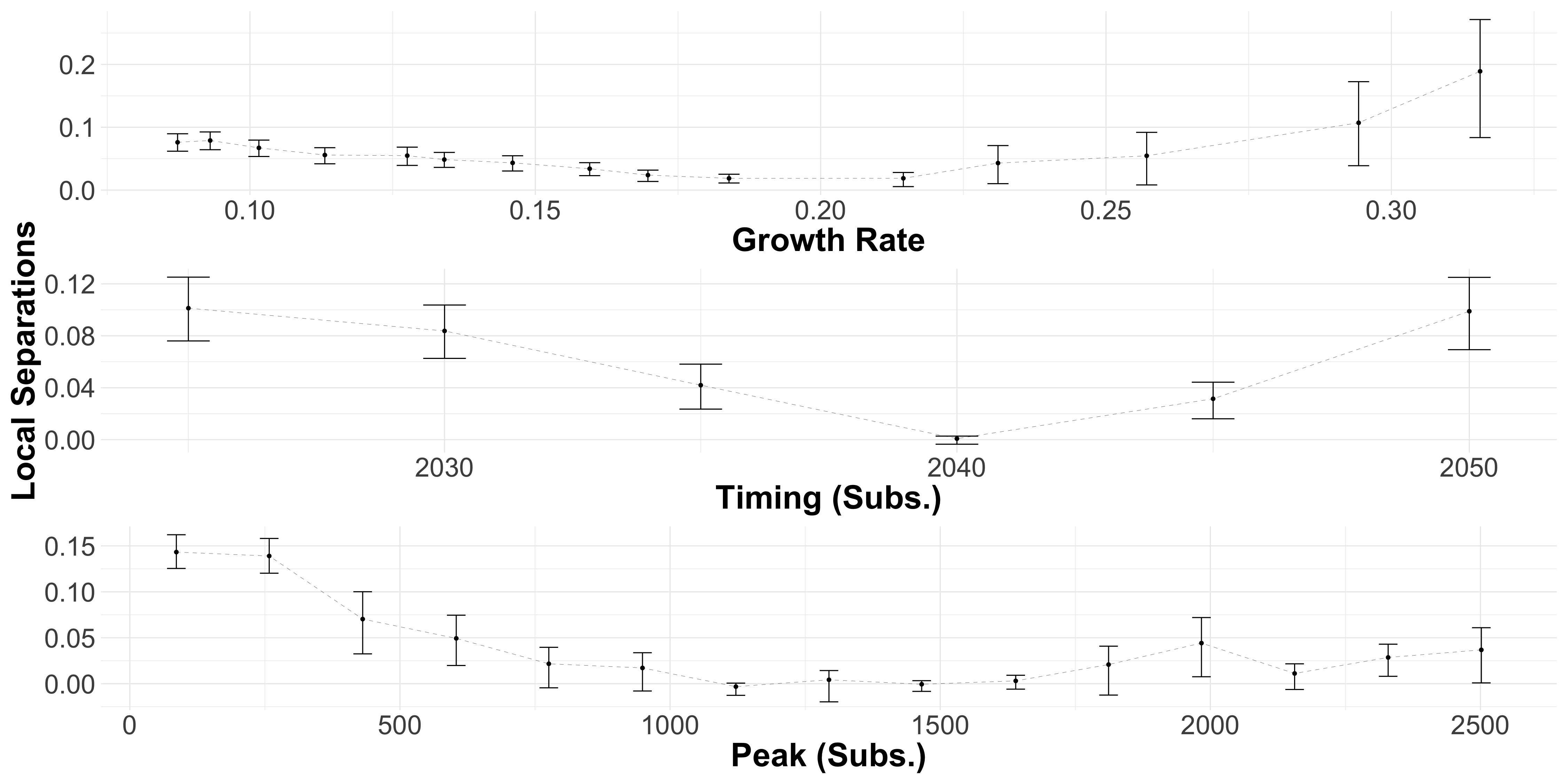}
    \caption{Local separations of three most important inputs for net removed CO\textsubscript{2} emissions in NDC scenario. In each plot, the x-axis represents the input, and the y-axis represents the estimated OT cost with bootstrap confidence intervals.}
    \label{fig:ndc_sep}
\end{figure}

\begin{figure}[H]
    \centering
    \includegraphics[width=1\linewidth]{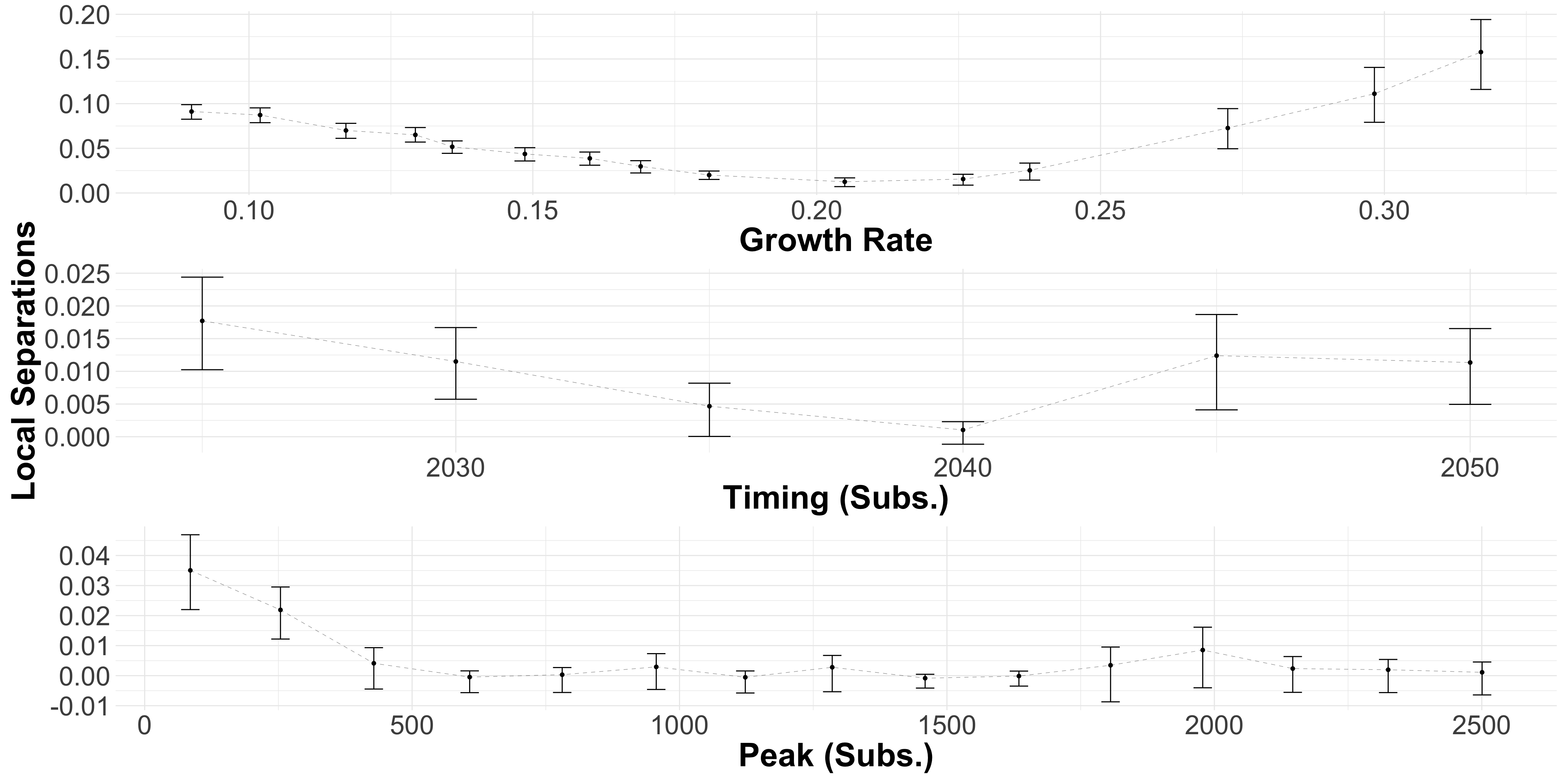}
    \caption{Local separations of three most important inputs for net removed CO\textsubscript{2} emissions in LTS scenario. In each plot, the x-axis represents the input, and the y-axis represents the estimated OT cost with bootstrap confidence intervals.}
    \label{fig:lts_sep}
\end{figure}

\begin{figure}[H]
    \centering
    \includegraphics[width=1\linewidth]{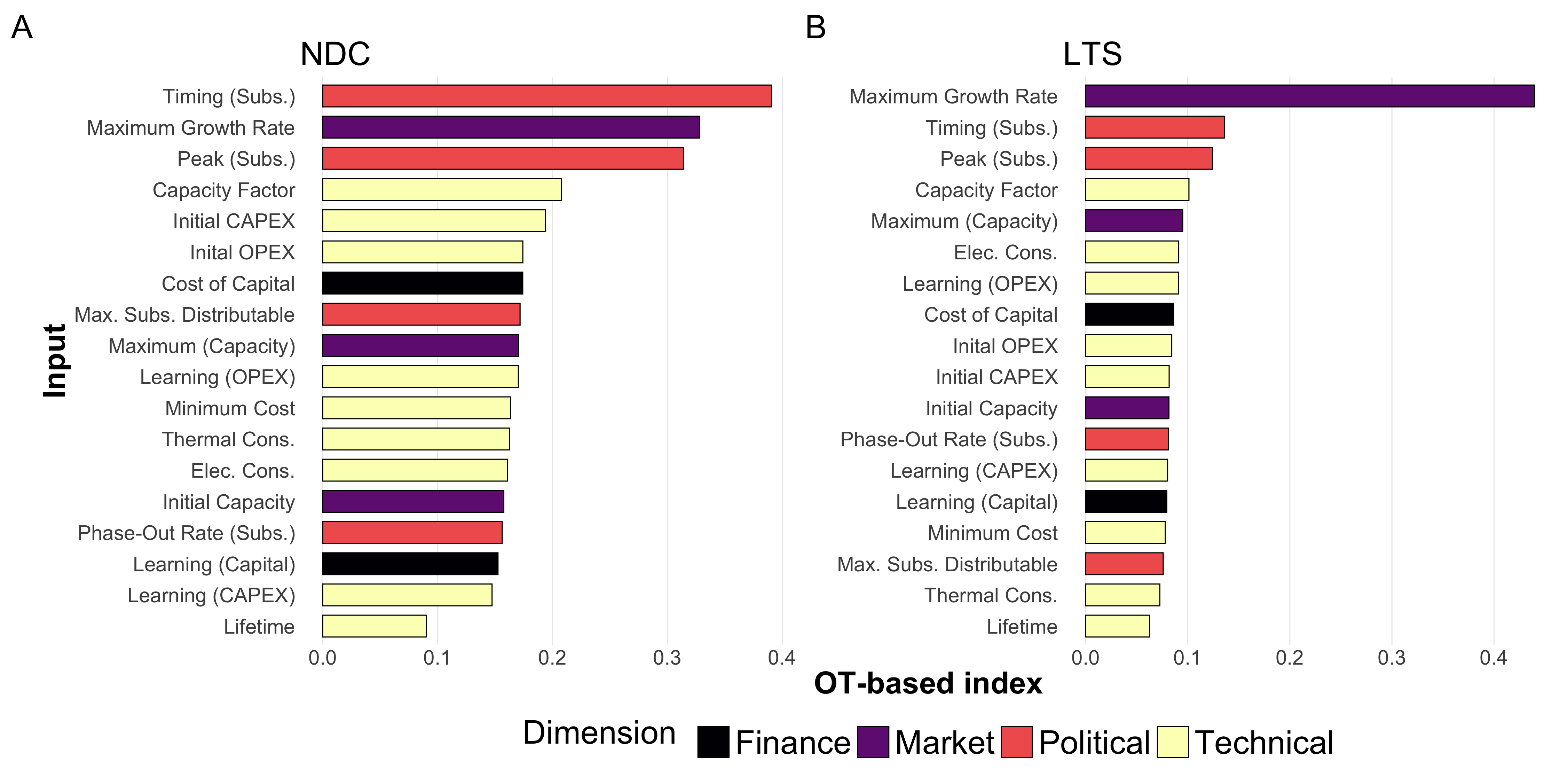}
    \caption{Sensitivity analyses for net removed CO\textsubscript{2} emissions in the two scenarios. The x-axis represents the OT-based index. On y-axis, each bar represents the input. Error bars are the 95\% bootstrapped confidence intervals. Colours represent the dimension of uncertainty. Inputs are ranked in descending order. Where inputs are differentiated by technology, we consider the maximum OT-based index among them.}
    \label{fig:ot_indices_filtered}
\end{figure}

\begin{figure}[H]
    \centering
    \includegraphics[width=1\linewidth]{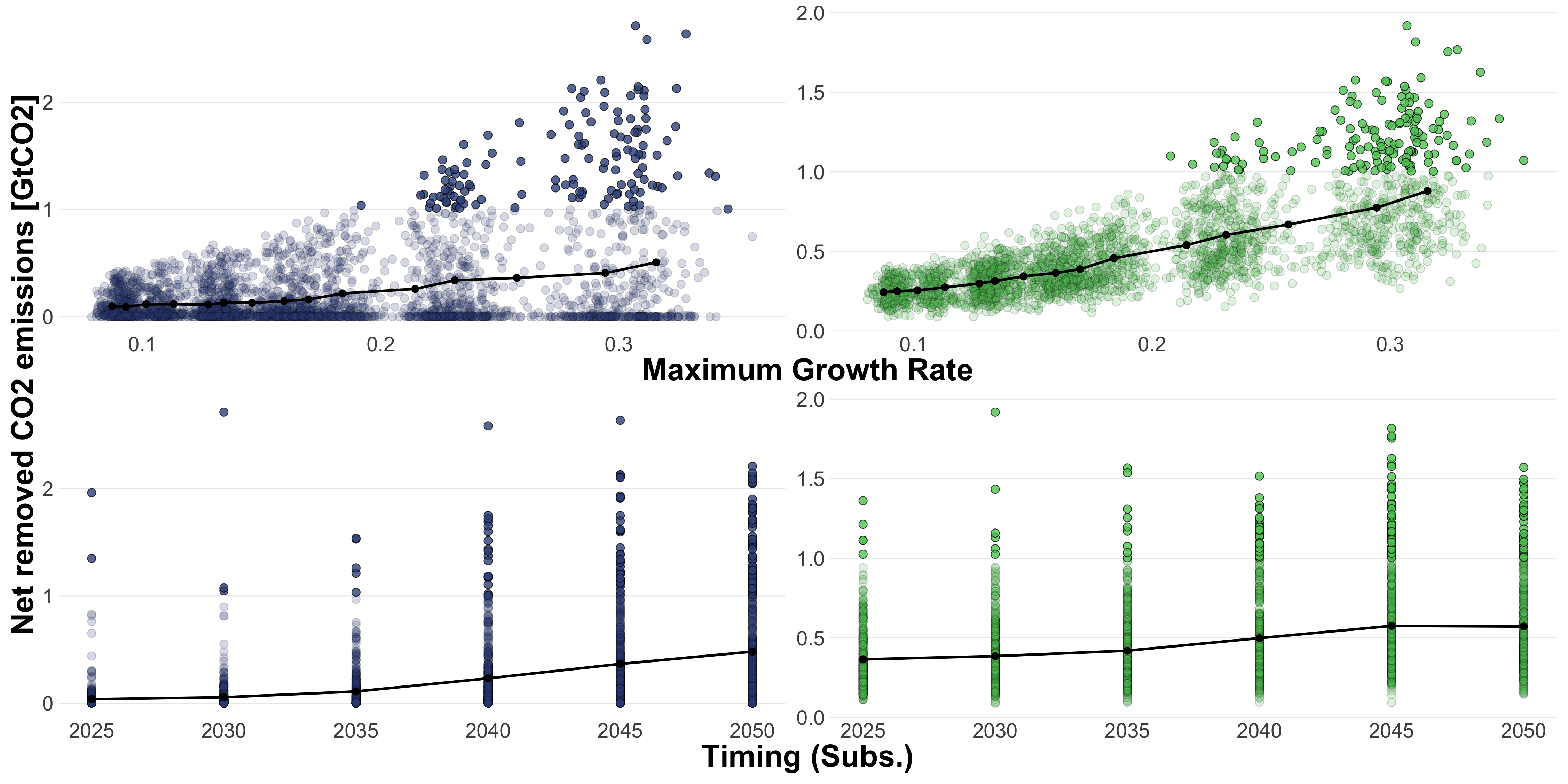}
    \caption{Partial dependency plots in the NDC scenario of the three most relevant inputs. The x-axis represents the value of one specific input, and the y-axis represents the corresponding net removed CO\textsubscript{2} emissions in 2050. The transparency is the 1 GtCO\textsubscript{2} threshold, and the black line represents the estimated conditioned mean.}
    \label{fig:enter-label}
\end{figure}

\section{Additional Results on Policy Gains}

\begin{figure}[H]
    \centering
    \includegraphics[width=1\linewidth]{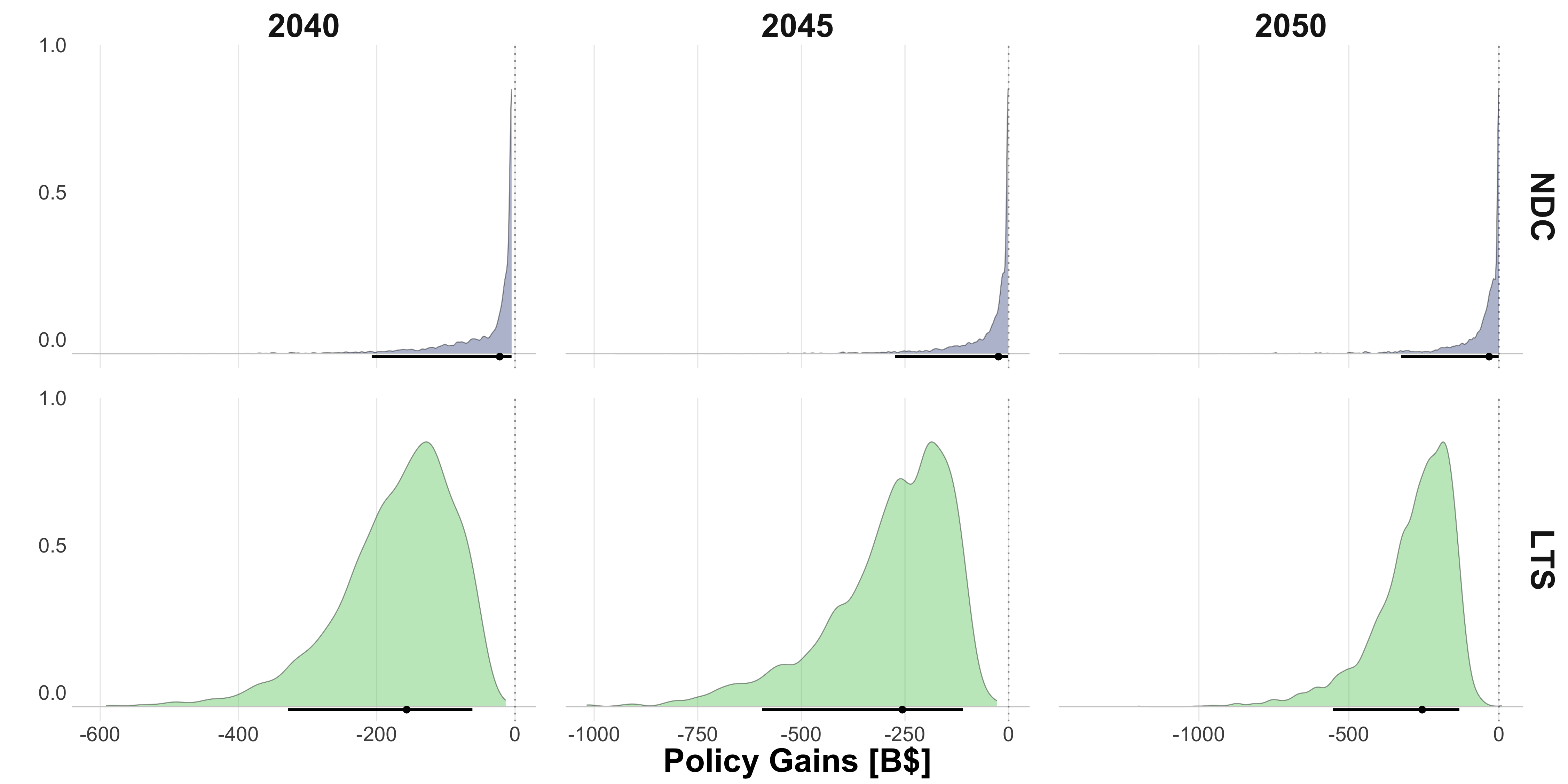}
    \caption{Distribution of yearly policy gains as variation of consumption from baseline. Each plot represents the empirical probability density. Transparency identifies the zero threshold. The black bars below the plots represent medians and 5th-95th quantile ranges.}
    \label{fig:enter-label}
\end{figure}

\begin{figure}[H]
    \centering
    \includegraphics[width=0.5\linewidth]{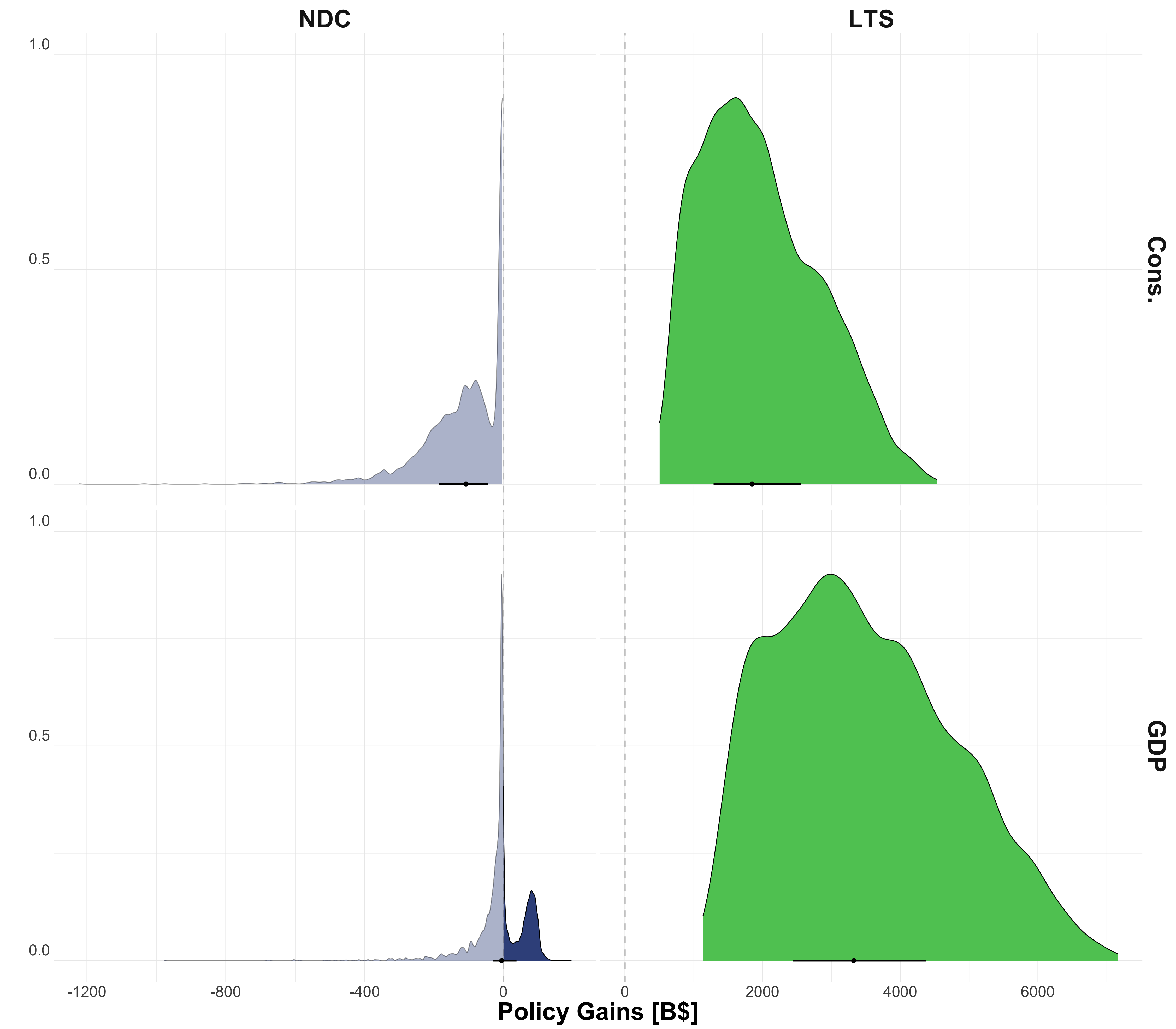}
    \caption{Policy gains in 2075 as variation of consumption (first row) and GDP (second row) from baseline in each scenario.}
    \label{fig:enter-label}
\end{figure}

\begin{figure}[H]
    \centering
    \includegraphics[width=0.5\linewidth]{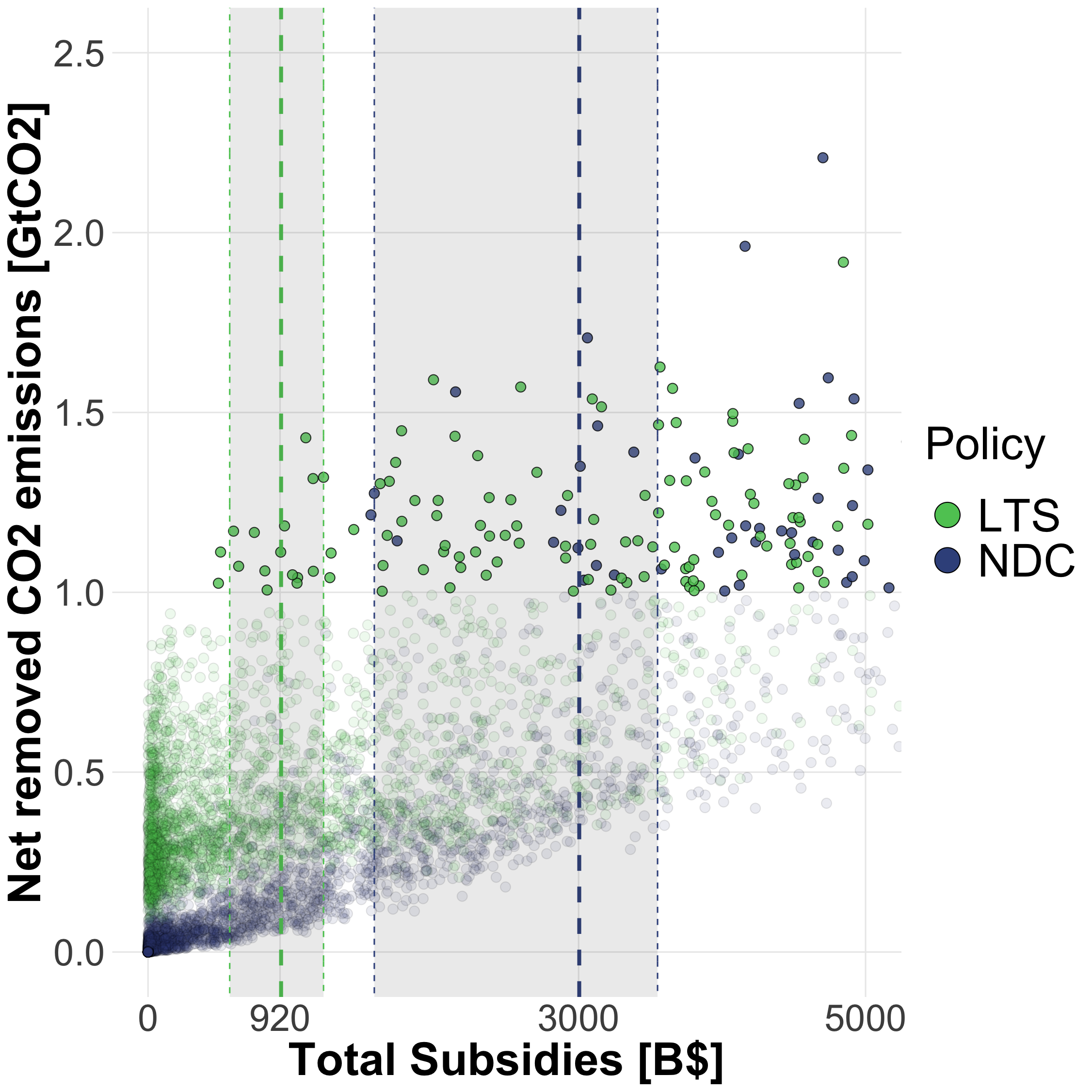}
    \caption{Relationship between net present value (3\% discount rate) of the subsidies distributed between 2025 and 2050 and installed capacity in 2050. The colored vertical lines represent the 5th percentile of the distribution of the points above 1GtCO\textsubscript{2}, and the shaded area is the 95\% bootstrap confidence interval. The transparency identifies the threshold of 1 GtCO\textsubscript{2}.}
    \label{fig:enter-label}
\end{figure}

\bibliography{sn-bibliography}